\documentclass[adp,fleqn]{w-art}
\usepackage{times}
\usepackage{w-thm}
\theoremstyle{plain}

\theoremstyle{definition}

\usepackage[]{graphicx}
\chardef\bslash=`\\ % p. 424, TeXbook

\begin{document}
%%    The information for the title page will be placed between
%%    \begin{document} and \maketitle. The order of most entries
%%    is determined by the class file and can not be changed by
%%    rearranging them. The maketitle command follows after the
%%    absract.
%%
%%    Most of the following commands will be completed by the publisher.
%%
%%    The copyrightyear is defined in the .clo file as the first argument
%%    of the copyrightinfo command. If the copyrightyear differs from that
%%    value it might be adjusted by the following definition:
%%
%% \renewcommand{\copyrightyear}{2002}% uncomment to change the copyrightyear.
%%
\DOIsuffix{theDOIsuffix}
%%
%% issueinfo for header and copyright line
\Volume{12}
\Issue{1}
\Copyrightissue{01}
\Month{01}
\Year{2003}
%%
%%    First and last pagenumber of the article. If the option
%%    'autolastpage' is set (default) the second argument may be left empty.
\pagespan{3}{}
%%
%%    Dates will be filled in by the publisher. The 'reviseddate' and
%%    'dateposted' (Published online) entry may be left empty.
\Receiveddate{1 September 2003}
\Reviseddate{10 December 2003}
\Accepteddate{23 December 2003 by U. Eckern}
%%\Dateposted{3 December 2002}
%%

%% \pretitle{Editor's Choice}

%% We have a short and a long form for the title. The short form
%% (optional argument) goes into the running head.

\title[Unconventional superconductivity and magnetism 
in Sr$_2$RuO$_4$]
{Unconventional superconductivity and magnetism 
in Sr$_2$RuO$_4$ and related materials}

%% Please do not enter footnotes or \inst{}-notes into the optional
%% argument of the author command. The optional argument will go into
%% the header.  If there is only one address the marker \inst{x} may be
%% omitted.

%% Information for the first author.
\author[I. Eremin et al.]{I. Eremin\footnote{Corresponding
     author: E-mail: {\sf ieremin@physik.fu-berlin.de}, 
Phone: +49 30 838 51422
Fax: +49 30 838 57422}\inst{1,2}} 
\address[\inst{1}]{Institut f\"ur Theoretische Physik, 
Freie Universit\"at Berlin, 14195 Berlin, Germany}
\address[\inst{2}]{Physics Department, 
Kazan State University, 420008 Kazan, Russian Federation}
\author[]{D. Manske
%\footnote{Second author footnote.}
\inst{1}}
\author[]{S. G. Ovchinnikov 
%\footnote{Third author footnote.}
\inst{3}}
\address[\inst{3}]{L.V. Kirensky Institute of Physics, 
Krasnoyarsk State University, Krasnoyarsk, Russia}
\author[]{J. F. Annett 
%\footnote{Third author footnote.}
\inst{4}}
\address[\inst{4}]{H.H.Wills Physics Laboratory, University of Bristol, 
Tyndall Ave, BS8-1TL, United Kingdom}
%%
%%    Information for the second author
%%\author[D. Manske]{D. Manske
%\footnote{Second author footnote.}
%%\inst{1}}
%\address[\inst{1}]{Second address}
%%
%%
%%    \dedicatory{This is a dedicatory.}
\begin{abstract}
We review the normal and superconducting state properties of the 
unconventional triplet superconductor Sr$_2$RuO$_4$ with an emphasis 
on the analysis of the magnetic susceptibility and the role played by 
strong electronic correlations. In particular, we show
that the magnetic activity arises from the itinerant electrons in the 
Ru $d$-orbitals and a strong
magnetic anisotropy occurs ($\chi^{+-} < \chi^{zz}$) due to
spin-orbit coupling. The latter results mainly
from different values of the $g$-factor for the transverse and
longitudinal components of the spin susceptibility (i.e. the matrix
elements differ). Most importantly, this anisotropy and the 
presence of incommensurate
antiferromagnetic and ferromagnetic fluctuations
have strong consequences for the symmetry of the superconducting 
order parameter. In particular, reviewing 
spin fluctuation-induced Cooper-pairing scenario in application to 
Sr$_2$RuO$_4$ we show how 
$p$-wave Cooper-pairing with line nodes between neighboring
RuO$_2$-planes may occur.

We also discuss the open issues in Sr$_2$RuO$_4$ like 
the influence of magnetic and non-magnetic 
impurities on the superconducting and normal state of Sr$_2$RuO$_4$.
It is clear that the physics of triplet superconductivity in Sr$_2$RuO$_4$ 
is still far from being understood completely 
and remains to be analyzed more in more detail.
It is of interest to apply the theory also to
superconductivity in heavy-fermion systems exhibiting spin fluctuations.
\end{abstract}

%% maketitle must follow the abstract.
\maketitle                   % Produces the title.

\tableofcontents
%% If there is not enough space inside the running head
%% for all authors including the title you may provide
%% the leftmark in one of the following three forms:

%% \renewcommand{\leftmark}
%% {First Author: A Short Title}

%% \renewcommand{\leftmark}
%% {First Author and Second Author: A Short Title}

%% \renewcommand{\leftmark}
%% {First Author et al.: A Short Title}

%% \tableofcontents  % Produces the table of contents.
\section{Introduction}

The phenomenon of superconductivity remains one of the most interesting 
problems of condensed matter physics. In particular, in recent years 
the material science development has revealed 
several interesting systems where 
high transition temperature superconductivity was found, in particular  
the families of high-$T_c$ cuprates \cite{bednorz} with a maximum 
$T_c$ of about 155 K. Soon after their discovery 
it was realized that the essential physics 
of cuprates takes place in the CuO$_2$-planes which is believed to be 
responsible for the high transition temperature. 
\begin{figure}[b]
\begin{center}
\includegraphics[width=0.4\textwidth]{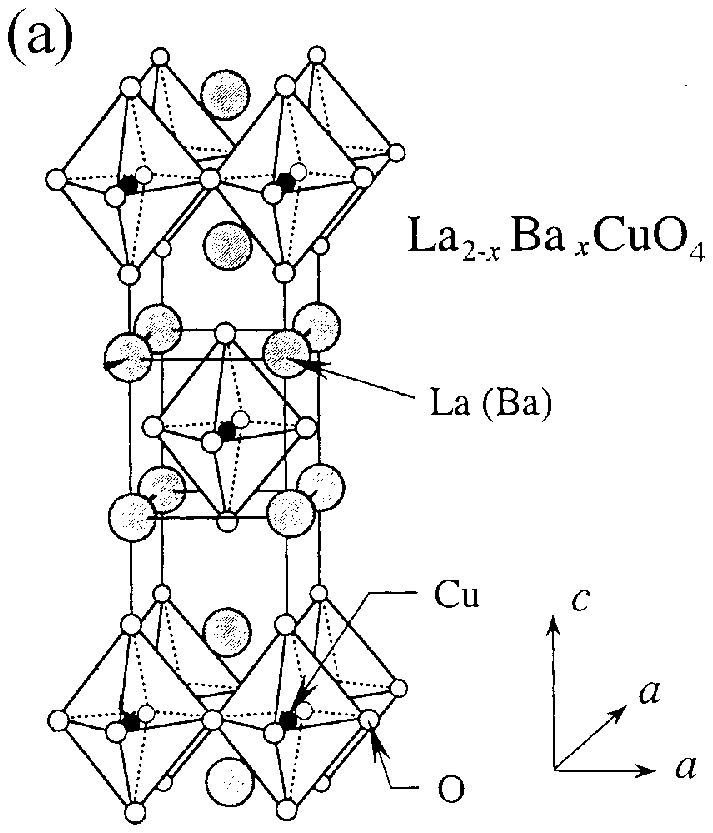}
\includegraphics[width=0.3\textwidth]{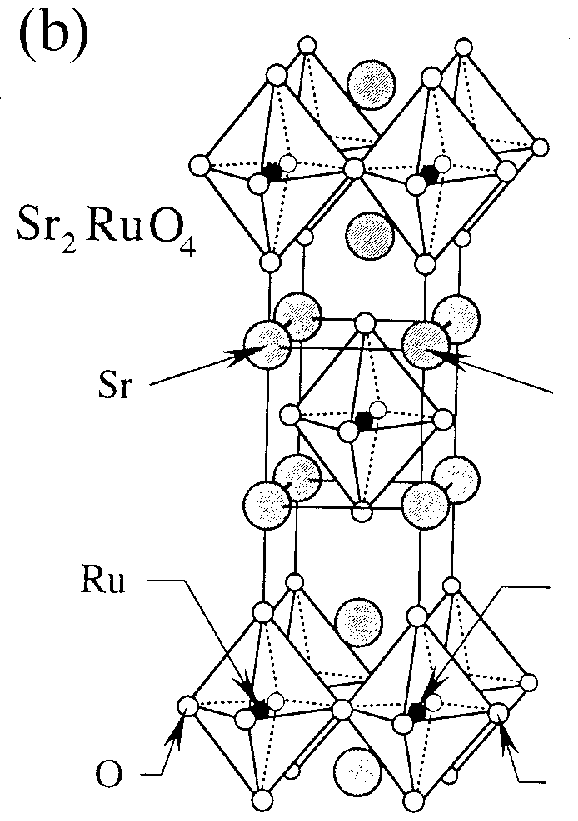}
\end{center}
\caption[]{(a) Crystal structures of (a) layered cuprate singlet 
superconductor
La$_{2-x}$Sr$_x$CuO$_4$ and (b) layered perovskite structure of the
triplet superconductor Sr$_2$RuO$_4$. Both structures are identical if 
La (Ba) is replaced by Sr and the CuO$_2$-plane is replaced by RuO$_2$-plane.}
\label{fig1}
\end{figure}

\begin{figure}
\begin{center}
\includegraphics[width=0.5\textwidth]{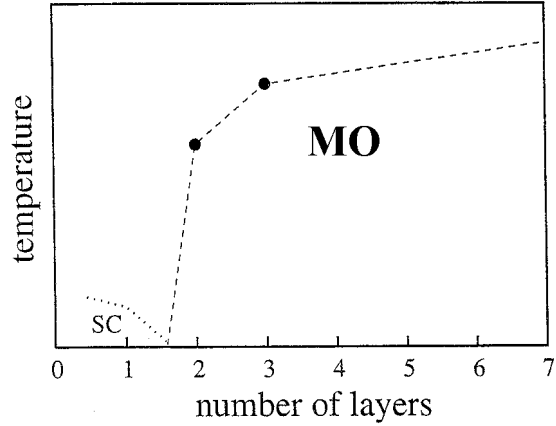}
\end{center}
\caption[]{Schematic phase diagram of
the ferromagnetic and superconducting systems (Rodllesen-Popper
crystals) Sr$_{n+1}$Ru$_n$O$_{3n+1}$, taken from Ref.
\protect\cite{sigri}. The number of layers is the parameter which
determines the transition between the two phases: MO=magnetically ordered 
and SC=superconducting. This phase digram suggests ferromagnetic 
excitations in the normal state of Sr$_2$RuO$_4$.}
\label{phasesr}
\end{figure}

In this connection the discovery of 
superconductivity in Sr$_2$RuO$_4$ with $T_c = 1.5$ K \cite{maeno} is of 
particular interest for several reasons. First, the crystal structure of 
Sr$_2$RuO$_4$ 
is identical to that of the parent compound of the high-$T_c$ superconductor 
La$_2$CuO$_4$ (see Fig. \ref{fig1} for illustration).
Both kinds of materials are highly two-dimensional and as Fig. \ref{fig1} 
shows the structure is almost identical to that of the La$_{2-x}$Sr$_x$CuO$_4$ 
superconductors. Both materials are oxides with conduction occurring 
in partially filled $d-$bands that are strongly hybridized with the oxygen 
$p$-orbitals.
Therefore, it was generally 
believed that a comparison of the normal and superconducting properties 
of the cuprates and Sr$_2$RuO$_4$ will give a deeper understanding of the 
nature of the high-$T_c$ in the cuprates. However, it has been found that the 
differences between Sr$_2$RuO$_4$ and the cuprates are larger than their 
general similarities might suggest. In Sr$_2$RuO$_4$ superconductivity 
occurs only at low temperatures and the normal state is a well-defined 
Fermi-liquid. This contrasts strongly with the anomalous normal state 
of the cuprates. Furthermore, it was soon found that the 
superconductivity in Sr$_2$RuO$_4$ is very interesting on its own.
In particular, there are clear indications that the superconducting 
state is unconventional. For example, the transition temperature is highly 
sensitive to impurities \cite{mac} and nuclear quadrupole 
resonance(NQR) experiments do not show a Hebel-Slichter peak in the 
spin-lattice relaxation at $T_c$ \cite{imad}. 
Shortly after the discovery of Sr$_2$RuO$_4$ 
it was suggested  that superconductivity might arise from odd-parity 
(spin-triplet) Cooper-pairs with total spin $S=1$ and a non-zero 
angular momentum which is reminiscent of the phases of superfluid 
$^3$He \cite{manfred}. The basis for this suggestion was the presence 
of ferromagnetism in the 
related compounds as SrRuO$_3$ and thus the expectation 
of ferromagnetic fluctuations in metallic Sr$_2$RuO$_4$. 
To support this picture the model phase diagram shown in Fig. \ref{phasesr} 
has been suggested. 
Here, one plots the phase of the
ferromagnetic and superconducting members of the so-called Rodllesen-Popper
series (Sr$_{n+1}$Ru$_n$O$_{3n+1}$) as a function of the numbers of
RuO$_2$-layers per unit cell, $n$. The infinite layer (SrRuO$_3$) is a
ferromagnet with $T_{Curie} \approx 165 $ K. For $n$=3 one finds
$T_{Curie} \approx 148$ K and for $n$=2 the substance orders magnetically at 
$T_{Curie} \approx 102$ K. This
demonstrates the tendency that $T_{Curie}$ is reduced with decreasing
layer number $n$ and suggests that even for $n$=1, when
supeconductivity occurs, one expects significant ferromagnetic
fluctuations which may play an important role for triplet
superconductivity in Sr$_2$RuO$_4$.

\begin{figure}
\begin{center}
\includegraphics[width=0.5\textwidth]{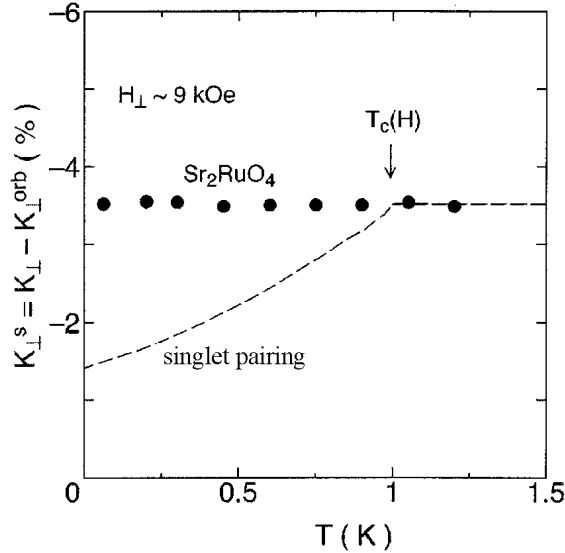}
\end{center}
\caption[]{Results for the uniform spin susceptibility, $\chi({\bf
q}=0,\omega=0)$ in the superconducting state of Sr$_2$RuO$_4$
as measured by NMR
Knight shift \protect\cite{imad}. One clearly sees that in
contrast to the singlet Cooper-pairing (illustrated by the dashed curve) 
where spin
susceptibility decreases upon cooling, the Knight shift in
Sr$_2$RuO$_4$ is unchanged by lowering $T$ below $T_c$. The Knight
shift doe not decrease below $T_c$, since the polarization induced by
the external magnetic field does not change in the superconducting
state in case of spin-triplet Cooper-pairs.}
\label{triplet}
\end{figure}

Meanwhile, a number of experiments indeed point 
towards spin-triplet Cooper-pairing. The most convincing evidence 
comes from the $^{17}$O NMR Knight shift data which shows that the spin 
susceptibility is not affected by the superconducting state for a magnetic 
field parallel to the RuO$_2$-plane \cite{imad}. In 
Fig. \ref{triplet} we show the corresponding 
experimental results. In conventional
superconductors the spin part of the Knight shift measured by NMR
decreases rapidly below $T_c$ due to the formation of singlet
Cooper-pairs. On the other hand, in a triplet superconductor with S=1
the spin part of the Knight shift should not change below $T_c$ for some field 
orientations, since
the polarization induced by the weak external magnetic field and
probed by NMR does not change. This behavior was observed in
Sr$_2$RuO$_4$ by Ishida {\it et al.} \cite{imad} and provides strong 
evidence for triplet Cooper-pairing.  

However, recently it has became clear that the situation is not that 
simple. 
For example, the intensive studies by means of inelastic neutron 
scattering (INS) reveal the presence of strong 
two-dimensional antiferromagnetic 
spin fluctuations at {\bf Q}$=(2\pi/3,2\pi/3)$ and no sign of  
ferromagnetic fluctuations in the normal state of Sr$_2$RuO$_4$ \cite{ins}. 
This fact, at first glance, cannot be compatible with triplet 
Cooper-pairing and thus the role of the antiferromagnetic fluctuations  
in the formation of superconductivity has to be clarified. 
Moreover, further analysis of the NMR data indicated the 
magnetic response to be strongly anisotropic \cite{ishida}. In particular,
the measured longitudinal component of the spin susceptibility $\chi_{zz}$ is 
much larger than the transverse one $\chi_{+-}$, while from an isotropic model
one expects $\chi_{+-}=2\chi_{zz}$. This anisotropy increases 
with decreasing temperature and reaches a maximum close to $T_c$ indicating 
its influence on the superconducting properties. 
Therefore, in view of the importance of spin fluctuations 
for the unconventional 
superconductivity in Sr$_2$RuO$_4$ their behavior should be understood in 
more detail. 

Another interesting question concerns the symmetry of the superconducting 
order parameter. While the analogy to the case of 
superfluid $^3$He leads to the suggestion of 
nodeless $p$-wave superconductivity in Sr$_2$RuO$_4$, 
recent analysis of the specific heat data \cite{maki} indicates a 
more complicated and anisotropic behavior of the superconducting order 
parameter in the first Brillouin Zone (BZ). Therefore, the physics of 
Sr$_2$RuO$_4$ is far from being completely understood.

Note, an extensive review on the experimental situation can be found in 
Ref. \cite{maemac}. In contrast to that review, 
in the present review we mainly analyze the role played by 
strong electronic correlations and spin fluctuations in Sr$_2$RuO$_4$. 
So far their role in 
determining the superconducting and normal state properties has not been 
addressed in detail. Probably, this was due to the success of the 
conventional band theory in explaining the various properties of Sr$_2$RuO$_4$.
However, its success is mainly due to the correct description of the 
Fermi surface topology 
in this compound. 
The Fermi surface shape, however, is hardly affected
by the electronic correlations, while on the other hand the 
bandwidth and the energy dispersion are strongly modified.
We will show that 
this is indeed the case for the ruthenates. Furthermore, we review the 
spin-fluctuation mediated Cooper-pairing scenario 
in application to Sr$_2$RuO$_4$ and 
address the important role of spin-orbit coupling in this system. 
In particular, we discuss the symmetry of the relevant superconducting order 
parameter within this scenario and, finally, investigate the role of 
impurities in Sr$_2$RuO$_4$.

\section{Superconductivity and magnetism 
in Sr$_2$RuO$_4$}

\subsection{Electronic structure and Fermi surface}

\begin{figure}
\vspace*{2ex}
\begin{center}
\includegraphics[width=0.4\textwidth]{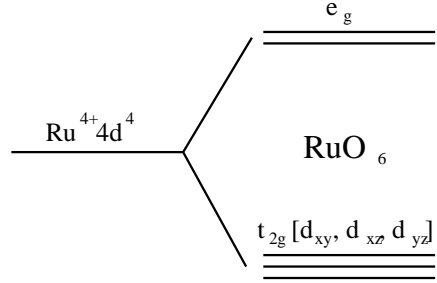}
\end{center}
\caption[]{Electronic structure of Sr$_2$RuO$_4$. The Ru$^{4+}$ ion 
corresponds to a 4d$^4$ level. The splitting of the $e_g$ and
$t_{2g}$ subshells is due to the RuO$_6$ crystal field. The orbitals
d$_{xy}$, d$_{xz}$, and d$_{yz}$ cross the Fermi level and form the $\alpha$-, 
$\beta$-, and $\gamma$-band. The resulting band structure is 
quasi-two-dimensional. Magnetic
activity results from the $t_{2g}$ subshell.}
\label{rutenstr0}
\end{figure}

In Sr$_2$RuO$_4$ the formal valence of the ruthenium ion is Ru$^{4+}$.
This leaves four electrons remaining in the $4d$-shell. Furthermore, the
Ru ion sits at the center of a RuO$_6$-octahedron and the crystal
field of the O$^{2-}$ ions splits the five $4d$-states into threefold
($t_{2g}$) and twofold ($e_g$) subshells as illustrated in Fig.
\ref{rutenstr0}. The negative charge of the $O^{2-}$ ions causes the
$t_{2g}$ subshell to lie lower in energy. Note, that these orbitals ($xz$,
$yz$, and $xy$) have lobes that point between the $O^{2-}$ ions lying
along the $x$, $y$, and $z$-axes. Electrons of these orbitals form the
Fermi surface (the so-called $\alpha$, $\beta$,
and $\gamma$-bands).  We assume that the most important
interaction of the carriers is with spin excitations described by the
spin susceptibility $\chi({\bf q},\omega)$ and resulting from Ru
$\alpha$-, $\beta$-, and $\gamma$-states.
Then, we have 3 bands formed out of the oxygen $p-d$-$\pi$ 
hybridized orbitals  and with a 
filling factor of 4/3. Since $p-d$-$\pi$ bonding
is weaker than the $p-d$-$\sigma$ realized in cuprates, the admixture of the 
oxygen orbitals is smaller in ruthenates than in cuprates. For example,
the contribution of the oxygen $p$-states close to the Fermi level is
only 16 percent in Sr$_2$RuO$_4$. At the same time the contribution of the 
Ru 4$d$-orbitals is about 84 percent \cite{noce}. LDA calculations
\cite{oguchi,singh} indeed confirm the existence of the three bands crossing 
the Fermi level. 

The band structure of Sr$_2$RuO$_4$ is quasi-two-dimensional 
and the electronic dispersion along the $c$-axis is very small \cite{singh}. 
\begin{figure}[h]
\begin{center}
\includegraphics[width=0.5\textwidth,angle=0]{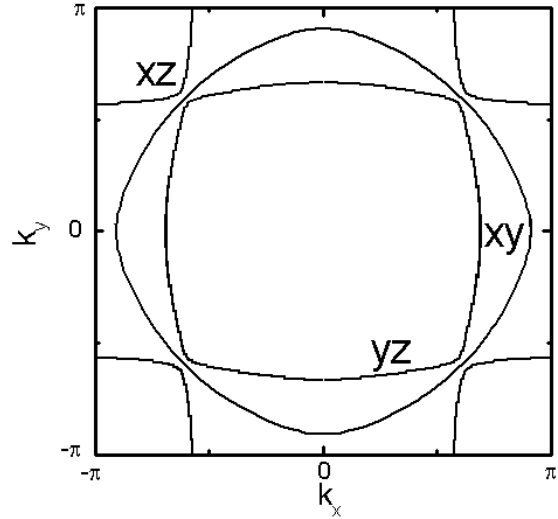}
\end{center}
\caption{ Calculated two-dimensional Fermi surface 
 of Sr$_2$RuO$_4$ using Eq. (1). The
 Fermi surface 
 consists of three sheets: two with electron-like topology and one
 with hole-like topology. Note, the quasi-one-dimensional character of the 
 $\alpha$(xz)- and $\beta$(yz)-bands. 
}
\label{fermruten}
\end{figure}
This is because the 
highly planar structure of Sr$_2$RuO$_4$ prevents substantial
energy dispersion along the $z$-direction, due to the large interplanar
separation of the RuO$_6$ octahedra. At the same time, in the
$ab$-plane neighboring RuO$_6$-octahedra share O ions which in turn
are $\pi$-bonded to the Ru ions. Thus, the $xy$-orbital will form a
two-dimensional band, while the $xz$- and $yz$-bands have only a
restricted one-dimensional character. 
Then, to describe the LDA results the following 
tight-binding energy dispersions can be introduced
\begin{equation}
\epsilon_{{\bf k}} = - \epsilon_0 -2t_x \cos k_x - 2t_y
\cos k_y +4t' \cos k_x \cos k_y
\quad .
\label{lind}
\end{equation}
The lattice constants $a=b$ have been set to unity. 
This model is a good parameterization of LDA calculations \cite{liebsch} 
if one chooses for $d_{xy}$-, $d_{zx}$-, and $d_{yz}$-orbitals the values
for the parameter set ($\epsilon_0, t_x, t_y, t'$) as (0.5, 0.42,
0.44, 0.14), (0.23, 0.31, 0.045, 0.01), and (0.23, 0.045, 0.31,
0.01) eV. The experimental investigation of the Fermi surface made by means 
of de-Haas-van-Alphen effect completely confirmed 
the LDA results \cite{macki}. The resulting Fermi surface is shown in 
Fig. \ref{fermruten}. Here, the $\alpha$-band is hole-like centered around 
$(\pi,\pi)$ point of the Brillouin Zone, while $\beta$ and $\gamma$-bands are 
electron-like ones centered around $\Gamma$-point. 
Furthermore, as one sees the $\alpha$ and $\beta$-bands 
are quasi-one-dimensional and their degeneracy is removed by the introduction 
of the interband (between $\alpha$ and $\beta$-bands) hopping 
$t_{\perp}$= 0.025 eV \cite{singh}.  

Despite of the large difference in the magnitudes of the resistivity 
in the RuO$_2$-plane and perpendicular to the planes, 
$\rho_c / \rho_{ab} \geq 500$,
the temperature dependences of $\rho_c$ and $\rho_{ab}$ for $T<25$ K both 
follow the Fermi-liquid behavior $ \propto T^2$ \cite{mac}. Above $T=25$ K 
small deviations from this law occur. At higher temperatures 
$\rho_{ab}$ becomes linear with a temperature dependence similar to the 
cuprates.

\subsection{ Dynamical spin susceptibility and electronic correlations}

As already mentioned in the introduction, the spin dynamics of Sr$_2$RuO$_4$ 
consists of ferromagnetic and incommensurate antiferromagnetic spin 
fluctuations at wave vector {\bf Q}$_i = (2\pi/3, 2\pi/3)$. Its 
role in the formation of triplet superconductivity in Sr$_2$RuO$_4$ 
requires special consideration.

Long-range magnetic order is absent in Sr$_2$RuO$_4$. On the other hand 
the uniform magnetic susceptibility is much larger than the Pauli 
susceptibility of the non-interacting electrons. Indeed the 'ab-initio' 
calculations of the Stoner factor $I N(0)$ (where $I$ is the exchange parameter
at ${\bf q}=0$, $N(0)$ is the density of states at the Fermi level) 
gives the following values $I N(0) = 0.82$ \cite{oguchi}, 
$I N(0) = 0.89$ \cite{singh}. Using a simple so-called Stoner criterium, 
this indicates that Sr$_2$RuO$_4$ is in 
the vicinity 
of the instability to ferromagnetism. 
In SrRuO$_3$ the corresponding 
Stoner parameter is $I N(0) = 1.23$ \cite{singh0}. Therefore, 
this system is a ferromagnetic metal  with magnetic moment $1.6 \mu_B$ and 
Curie temperature $T_c = 150$ K. The reason for the enhanced ferromagnetic 
exchange fluctuations in SrRuO$_3$ in comparison to Sr$_2$RuO$_4$ 
is the stronger 
$p-d$ hybridization and thus a larger contribution of the oxygen $p$-orbitals 
in the density of states at the Fermi level. 

However, what is interesting in Sr$_2$RuO$_4$ is not only the presence of 
ferromagnetic fluctuations at ${\bf q}=0$ (from the Stoner instability),  
but also the incommensurate 
antiferromagnetic fluctuations at the wave vector 
${\bf Q}_i = (2\pi/3, 2\pi/3)$ as observed in inelastic neutron scattering 
experiments \cite{ins}. What is the origin of these fluctuations?
In fact they originate from the nesting properties of the 
quasi-one-dimensional $xz$- and $yz$-bands. 
\begin{figure}[h]
\begin{center}
\includegraphics[width=0.6\textwidth,angle=0]{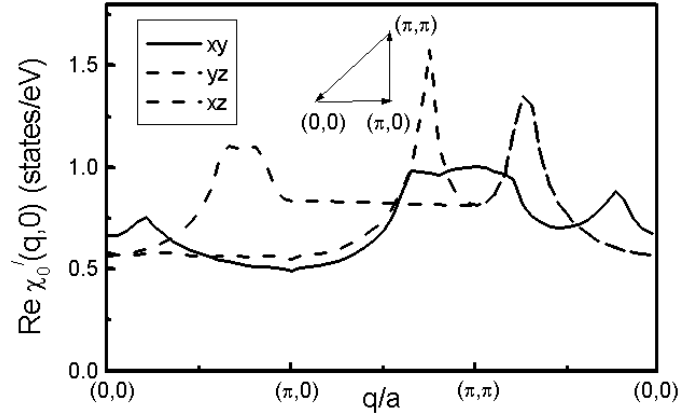}
\end{center}
\caption{ Calculated real part of the Lindhard spin susceptibility 
$\chi_0^{l}$ ($l$ refers to the band indices) in the
normal state of Sr$_2$RuO$_4$
along the route $(0,0)\to(\pi,0)\to(\pi,\pi)\to(0,0)$ in
the first BZ for the 
three different orbitals ($xz$,$yz$, and $xy$) crossing the
Fermi level. Due to the nesting of the $xz$ and $yz$-bands their
susceptibilities show an enhancement at the incommensurate
antiferromagnetic wave vector ${\bf Q}_i = (2\pi/3,2\pi/3)$.  The
response of the $xy$-band is more isotropic, but significantly larger
than in the normal metal due to the vicinity of the Van-Hove singularity.
}
\label{ruthiohne}
\end{figure}
In order to see this we show in Fig. \ref{ruthiohne} the results for the 
Lindhard response functions calculated for the different bands using their
tight-binding dispersions
\begin{equation}
\chi_0^{i}({\bf q},\omega)= - \frac1N \sum_{{\bf k}} 
\frac{f_{\bf k}^{i}-f_{\bf k+q}^{i}}
{\epsilon_{\bf k}^i - \epsilon_{\bf k+q}^i + \omega +i0^+} \quad ,
\end{equation}
where $i$ refers to the band index and $f_{\bf k}$ is the Fermi distribution 
function.  
As seen in Fig. \ref{ruthiohne}, due to the pronounced nesting of the $xz$ and
$yz$-bands their  susceptibilities display peaks at 
{\bf Q}$_i =(2\pi/3,2\pi/3)$, while the $xy$-band does not show any
significant feature. The response of the $xy$-band
is enhanced due to the presence of the van-Hove singularity close to the
Fermi level. Then, it becomes clear that the features observed by INS
relate mainly to the magnetic response of the $xz$- and $yz$-bands. 
Furthermore, as has been shown in Ref. \cite{singh1} that 
the Stoner enhancement 
related to the appearance of the incommensurate antiferromagnetic fluctuations
is even stronger than the ferromagnetic one and is of the order of 1.
This means that Sr$_2$RuO$_4$ is almost 
unstable with respect to the spin density wave formation (SDW). 
Note, that both ferromagnetic and antiferromagnetic fluctuations are seen 
 experimentally in the normal state of 
Sr$_2$RuO$_4$ by means of $^{17}$O Knight shift \cite{imai} and  
by inelastic neutron scattering \cite{ins}, respectively. 

These results suggest that 
an effective Hamiltonian for describing the physics in
Sr$_2$RuO$_4$ is a two-dimensional three-band Hubbard Hamiltonian
\begin{equation}
H=\sum_{{\bf k}, \sigma} \sum_{\alpha} \epsilon_{{\bf k} \alpha} 
a_{{\bf k}, \alpha \sigma}^{+} a_{{\bf k}, \alpha \sigma} +
 \sum_{i,\alpha} U_{\alpha} \, n_{i \alpha \uparrow}
n_{i \alpha \downarrow}
\quad ,
\label{hamiltsr}
\end{equation}
where $a_{{\bf k}, \alpha \sigma}$ is the Fourier transform of the
annihilation operator for the $d_{\alpha}$ orbital electrons ($\alpha
= xy, yz, zx$) and $U_{\alpha}$ is an effective on-site Coulomb
repulsion. The hopping integrals $t_{{\bf k}\alpha}$ denote the energy
dispersions of the tight-binding bands, see Eq.(\ref{lind}).

As a result the spin dynamics in Sr$_2$RuO$_4$ is determined by the 
competition 
and the delicate balance between ferromagnetic and incommensurate 
antiferromagnetic fluctuations. The natural question arises what is the 
influence of these magnetic fluctuations on the triplet superconductivity 
in Sr$_2$RuO$_4$? In the following subsection we will study the formation of triplet of 
superconductivity in Sr$_2$RuO$_4$ driven by the spin fluctuation exchange.

In the remainder of this section we discuss another aspect of the ruthenates, 
namely strong electronic correlations.
As a matter of fact the success of 'ab-initio' methods in describing magnetism
of Sr$_2$RuO$_4$ appear to invalidate, at first sight, the discussion of the 
role of strong electronic correlations. On the other hand, the effects of 
strong electronic correlations are small if the bandwidth 
of the conduction band $W$ is much larger than the on-site Coulomb 
repulsion, $U$. Given that the bandwidth is of the order of 1 eV 
there is no reason to think that $W>>U$ in Sr$_2$RuO$_4$. 
To estimate the value of $U$ one has to remember that for 4$d$-electrons 
 $U$ has to be smaller than for $3d$-electrons, since the average radius 
of the 4$d$-shell is approximately two times larger than for the 3$d$-shell. 
Then a simple estimates gives U$_{Ru} \approx 1$ eV. This means that 
Sr$_2$RuO$_4$ is in the regime of intermediate electronic correlations when 
$U \sim W$. 

As we mentioned earlier, despite of the importance of the 
electronic correlations in Sr$_2$RuO$_4$ the relative  
success of the band structure theory can be easily explained. The 
correct description of the ferromagnetic and antiferromagnetic fluctuations 
comes from the correct Fermi surface topology which is not affected by 
the electronic correlations. However, the bandwidth and the energy dispersion 
are strongly affected by the electronic correlations in Sr$_2$RuO$_4$. 
For example, it has been shown by Liebsch and Lichtenstein 
\cite{liebsch} that in order to explain 
the bandwidth as measured by angle-resolved photoemission 
spectroscopy (ARPES) the electronic correlations have to be taken into account.
In particular, using dynamical-mean-field theory (DMFT) they have shown that, 
while 
the Fermi surface topology is the same independent on electronic 
correlations, the bandwidth is different, namely the correlations reduce
the bandwidth of the all bands in Sr$_2$RuO$_4$. We will further address this 
issue discussion the phase diagram of Ca$_{2-x}$Sr$_x$RuO$_4$.

\subsection{Superconducting state of Sr$_2$RuO$_4$: role of 
spin fluctuations and symmetry of the superconducting order parameter}

Even before experimental results have indicated triplet superconductivity 
in Sr$_2$RuO$_4$, its possibility had been theoretically predicted by
Sigrist and Rice \cite{manfred}. In particular, they considered the triplet
superconductivity on a square lattice as an electronic analogy of the 
superfluid A-phase in $^3$He.

For the spin-triplet Cooper-pairing the wave function can be written 
as a matrix in spin space
\begin{eqnarray}
\Psi = g_1(k) | \uparrow \uparrow> + g_2(k) \left( | \uparrow \downarrow> 
+ | \downarrow \uparrow> \right) + g_3 (k)  | \downarrow \downarrow>
& = & \left( \begin{array}{cc}
g_1(k) & g_2(k) \\
g_2(k) &  g_3(k)
\end{array} \right) \quad,
\label{1}
\end{eqnarray}
where the eigenvalues of the S$_z$ operator with projection +1, 0, -1 have the
form
\begin{eqnarray}
| \uparrow \uparrow> = \left( \begin{array}{cc}
1 & 0 \\
0 & 0
\end{array} \right) \quad, \quad
| \uparrow \downarrow> + | \downarrow \uparrow> = \left( \begin{array}{cc}
0 & 1 \\
1 & 0
\end{array} \right) \quad, \quad
| \downarrow \downarrow> =  \left( \begin{array}{cc}
0 & 0 \\
0 & 1
\end{array} \right) \quad.
\label{2}
\end{eqnarray}
Another possibility to express the Cooper-pair wave function is to use the
basis of the symmetric matrices
\begin{equation}
i{\bf \sigma}\sigma_y = (i \sigma_x \sigma_y, i \sigma_y \sigma_y, 
i \sigma_z \sigma_y ).
\end{equation}
Then the Cooper-pair wave function has the form
\begin{eqnarray}
\Psi = i\left( {\bf d(k) \cdot \sigma} \right) = 
\left( 
\begin{array}{cc}
-d_x + id_y & d_z \\
d_z & d_x + id_y
\end{array}
\right) \quad.
\end{eqnarray}
The components of the vector {\bf d} can be expressed linearly via the
amplitudes of $g_{\alpha} ({\bf k})$: 
\begin{eqnarray}
g_1 = -d_x + id_y, \quad g_2 = d_z, \quad g_3 = d_x +id_y
\end{eqnarray}
Due to the 
Pauli principle the orbital part of the wave function with total $S=1$
has to be odd, i.e. the orbital quantum number $l=1, \, 3, \, ...$.
Therefore, one can speak about $p, \, f, \, ...$ Cooper-pairing in a 
continuum system. In particular, the
amplitudes g$_{\alpha} ({\bf k})$ and the vector {\bf d(k)} have to be 
the odd functions of the momentum {\bf k}.

In the tetragonal crystal structure such as that of Sr$_2$RuO$_4$,
the order parameter {\bf d(k)} has to belong to the corresponding 
irreducible representation 
of the D$_{4h}$ group. There are several representations
\cite{manfred, annett} and 
the proper choice can be made by comparison to experiment. 
For example, the spin state of the Cooper-pair wave
function can be obtained from the measuring the Knight shift in NMR  
and polarized inelastic neutron scattering experiments below $T_c$. 
The orbital state can be  measured by muon spin 
relaxation ($\mu$SR) indicating time-reversal symmetry breaking. The square 
symmetry of the vortex lattice can also be interpreted as evidence for 
$p$-wave pairing.

As we have shown in the introduction, the NMR Knight shift experiments indicate 
spin-triplet Cooper-pairing in Sr$_2$RuO$_4$ \cite{imad}. The same 
conclusion has been drawn from the polarized neutron scattering
studies \cite{60}. The time-reversal symmetry breaking detected by means of the
$\mu$SR \cite{61} suggests the superconducting order parameter $E_u$ with $S_z
= 0$ and $l_z = \pm 1$ (so-called chiral $p$-wave state):
\begin{equation}
d_z = \hat{ \bf z}\Delta_0 (\sin k_x \pm i \sin k_y), 
\quad \hat{ \bf z}= |S_z = 0> \quad.
\end{equation}
The amplitude of this gap is isotropic, $\Delta_{\bf k} = 
\Delta_0 \sqrt{k_x^2 + k_y^2}$,  and does not have a node at the Fermi surface
of Sr$_2$RuO$_4$.

So far, the above analysis is quite general. Next, we assume 
the superconductivity is driven by the exchange of spin
fluctuations. 
The interesting question in the ruthenates is why does triplet
Cooper-pairing take place in the presence of strong incommensurate
antiferromagnetic spin fluctuations at ${\bf Q}_i = (2\pi/3,2\pi/3)$
originating from the $xz$- and $yz$-bands and only relatively weak
ferromagnetic fluctuations arising from the $xy$-band. In general, to
answer this question the BCS gap equation (see Eq. (\ref{gaprut})) must be
solved assuming different possible symmetries of the order parameter
and corresponding pairing interaction for singlet, triplet with $S_z =
0$ projection and triplet with $S_z = \pm 1$ projection
Cooper-pairing.
For the analysis of the interaction between quasiparticles and spin
fluctuations (ferromagnetic and incommensurate antiferromagnetic) we
use the generalized Eliashberg like theory for the interaction between
quasiparticles and spin fluctuations \cite{scalapino}.   
We extend this theory to become
a three-band theory, as necessary in Sr$_2$RuO$_4$ \cite{ueda,arita}.  

In the normal state both the self-energy and thermal Green's functions become
matrices of $3\times 3$ form, {\it i.e.} $G_{i,j,m} ({\bf
  k},\omega_n)$ and $\Sigma_{i,j,m} ({\bf k},\omega_n)$, where $i,j,m$
refer to the band indexes of the $xy, yz,$ and $xz$-orbitals. The
corresponding Dyson equation is given by
\begin{equation}
\left[\hat{G} ({\bf  k},\omega_n)\right]^{-1} =
\left[\hat{G}^0 ({\bf  k},\omega_n)\right]^{-1} - 
\hat{\Sigma} ({\bf  k},\omega_n),
\label{dyson3}
\end{equation}
where $\hat{G}^0_{i,j,m} ({\bf k},\omega_n)$ is the matrix of the bare
Green's function determined via the tight-binding energy dispersions
for the $xy$, $yz$, and $xz$-bands. The self-energy is given by
\begin{eqnarray}
\Sigma_{i,j,m} ({\bf  k},\omega_n) & = & T\sum_{q,l} V^1_{i,j,m} 
({\bf q}, \omega_l) \times 
G_{i,j,m} ({\bf  k-q},\omega_n-\omega_l) 
\quad,
\end{eqnarray}
where $V^1_{i,j,m} ({\bf q}, \omega_l)$ is an effective interaction
\begin{figure}[t]
\begin{center}
\includegraphics[width=0.7\textwidth,angle=-1]{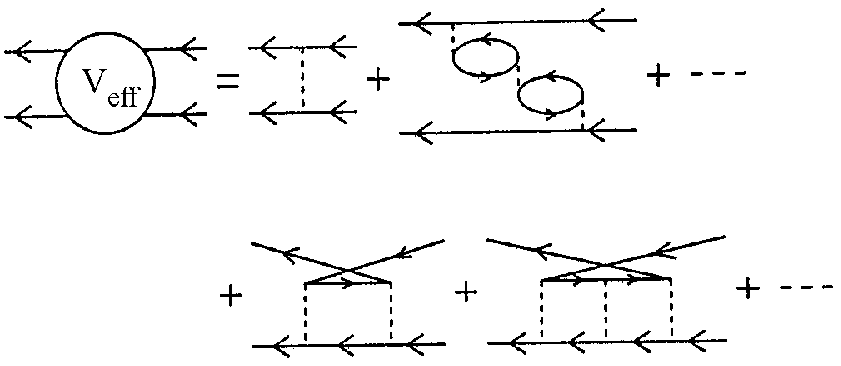}
\label{gamma}
\end{center}
\vspace{1ex}
\caption[]{Lowest order RPA diagrams for the effective pairing interaction
$V_{eff}$ for singlet pairing resulting from the
exchange of longitudinal and transverse spin and charge
fluctuations. The solid lines refer to the one-particle Green's
function and the dashed lines denote an effective Coulomb
interaction $U$. The first diagram leads to a renormalized
chemical potential. Note that for singlet pairing only an even
number of bubble diagrams occur due to Pauli's principle.}
\label{gammasf}
\end{figure}
between quasiparticles and spin fluctuations. As shown in Fig.7 
it consists of an infinite series of diagrams including charge
and spin fluctuations. This is similar to the case of cuprates, 
however, some important differences are
present in the random phase approximation.

\begin{figure}
\begin{center}
\includegraphics[width=0.7\textwidth]{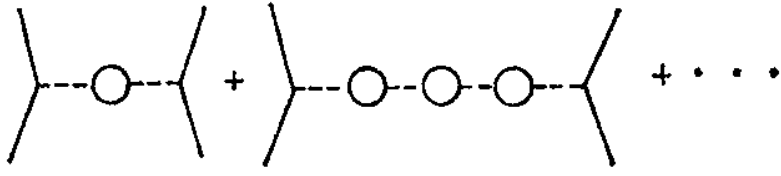}
\end{center}
\caption[]{Additional RPA diagrams of the lowest order 
for triplet Cooper-pairing
with odd number of bubbles that refer to longitudinal charge and
spin fluctuations.}
\label{tripltheory}
\end{figure}

Most importantly we must now consider the diagrams shown in Fig.
\ref{tripltheory} with an odd number of bubbles that contribute to the
triplet pairing. This is in contrast to singlet pairing in cuprates
where an even number of bubbles occur (see Fig. \ref{gammasf}).
Furthermore, the transverse (+-) and the longitudinal (zz) parts of the spin
susceptibility are different. The reason for this (as we show later) 
is the presence of spin-orbit
coupling. We discuss its role in detail in the next section.  
Then the effective pairing interaction in the
$3\times 3$ form including transverse and longitudinal spin
fluctuations and also charge fluctuations is given by
\begin{eqnarray}
V^1_{i,j,m} ({\bf q}, \omega_l) = \frac{1}{2}
V_{i,j,m}^{sp,zz}({\bf q},
\omega_l) + V_{i,j,m}^{sp,+-}({\bf q},\omega_l) +
\frac{1}{2} V^{ch}_{i,j,m} ({\bf q},\omega_l)
\quad,
\end{eqnarray}
where
\begin{eqnarray}
V^{sp,zz}  & = & U^2  \frac{\chi^{sp,zz}_0}{1-U\chi^{sp,zz}_0}, 
\,\,\,\,\,\,\,\,
V^{sp,+-} = U^2\frac{\chi^{sp,+-}_0}{1-U\chi^{sp,+-}_0} 
\end{eqnarray}
describe coupling to spin density fluctuations and
\begin{eqnarray}
V^{ch} &  = & U^2\frac{\chi^{ch}_{0}}{1+U\chi^{ch}_{0}}
\end{eqnarray}
to charge density fluctuations.  $\chi^{ch}_{0}$, $\chi^{sp,+-}_{0}$
are the irreducible parts of the charge and spin susceptibilities 
respectively. 
Note that in the Bethe-Salpether equation shown 
diagrammatically in Fig. \ref{gammasf}  for singlet Cooper-pairing 
it is necessary to have an 
even number of bubbles and include the ladder diagrams. In the case of 
triplet Cooper-pairing the contribution of the ladder diagrams is zero 
and an odd number bubbles diagrams occur due to Pauli's principle 
shown in Fig. \ref{tripltheory}. Since the Feynman rules require a 
$(-1)$ for each loop an extra minus sign enters the gap equation via 
$V_t^{eff}$. 
 
The magnetism in the ruthenates resulting from the quasiparticles in 
the $t_{2g}$-orbital is itinerant and thus the magnetic response is 
created by the same electrons that form also the Cooper-pairs.  Then, 
for example, the irreducible part of the charge susceptibility 
$\chi^{ch}_{0}$ is defined in terms of the electronic Green's 
functions 
\begin{equation} 
\chi^{ch}_{0}({\bf q})= 
\frac{1}{N}\sum_{\bf k} G({\bf  k+q}) G({\bf k}), 
\end{equation}  
where $G({\bf k})$ is the single-electron Green's function (we omit here 
the band indices for simplicity). The longitudinal and transverse 
components of the spin susceptibilities are also calculated in terms 
of the electronic Green's functions. In contrast to the cuprates, in the 
ruthenates $\chi^{sp,zz}_0$ and $\chi^{sp,+-}_0$ are different due to 
the magnetic anisotropy \cite{imad}. 
 
In the ruthenates the general matrix form of the superconducting gap 
equation for $\Delta({\bf k},\omega)$ is similar to the case of 
cuprates. To determine the superconducting transition temperature 
$T_c$ one must solve the following set of linearized gap equations 
to obtain $\lambda_{\mu}(T)$: 
\begin{eqnarray} 
\lambda_{\mu} \Delta_{\mu,l,m} ({\bf k},\omega_n) 
& = & 
-\frac{T}{N} \sum_{k',\omega_j} \sum_{l',m'} 
V_{\mu,l,m}^{(2)} ({\bf k-k'}, 
\omega_n-\omega_j) \, G_{ll'}({\bf k'},\omega_j) 
\nonumber\\ 
& & 
\times \, G_{mm'}(-{\bf k'},-\omega_j) 
\, \Delta_{\mu,l',m'} ({\bf k'},\omega_j). 
\label{gaprut} 
\end{eqnarray} 
Here, $\lambda_{mu}$ is an eigenvalue and $T_c$ is determined from the 
condition $\lambda(T_c)=1$. 
 
Note, the interband coupling will provide a single $T_c$ for all three 
bands.  The pairing potential $V^{(2)}$ (taken for singlet or triplet 
pairing) controls which state gives the lowest energy and determines 
singlet or triplet Cooper-pairing. Three possibilities may occur:\\[2ex] 
\hspace*{1cm}(a) singlet pairing, \\ 
\hspace*{1cm}(b) triplet pairing with the total spin S$_z = 0$ of 
the Cooper-pair \\ \hspace*{1.6cm}wave function, \\ 
\hspace*{1cm}(c) triplet pairing with the total 
spin S$_z = \pm 1$.\\[1ex] 
 
Note, $a$ $priori$ one cannot judge which pairing state is realized in 
the ruthenates due to the presence of both antiferromagnetic and 
ferromagnetic fluctuations.  Therefore, one has to solve the gap 
equations for all three possibilities. 
Thus, for singlet pairing we use 
\begin{equation} 
V_s^{(2)} = \frac{1}{2} V_{sp}^{zz} + V_{sp}^{+-} -\frac{1}{2}V_{ch} 
\quad. 
\end{equation} 
For triplet pairing with  S$_z = \pm 1$ we take  
\begin{equation} 
V_{tr1}^{(2)} = -\frac{1}{2} V_{sp}^{zz} -\frac{1}{2}V_{ch} 
\label{treplo} 
\end{equation} 
resulting from the set of diagrams with odd number of bubbles (see 
Fig. \ref{tripltheory}). 
Finally, for triplet pairing with $S_z = 0$ we get 
\begin{equation} 
V_{tr0}^{(2)} = \frac{1}{2} V_{sp}^{zz} - V_{sp}^{+-} 
-\frac{1}{2}V_{ch}  
\quad. 
\label{sedlo} 
\end{equation} 
Therefore due to magnetic anisotropy we have separated the longitudinal and 
transverse parts of the spin fluctuations.   
Note, transverse and longitudinal 
spin fluctuations contribute differently to the singlet and triplet 
Cooper-pairing.  In particular, in the spin-triplet case the 
longitudinal and transverse spin fluctuations do not act additively as in 
the case of singlet Cooper-pairing. As a result, a strong reduction of 
the superconducting transition temperature $T_c$ is expected. 
The effect of charge 
fluctuations is expected to be small in the superconductivity in 
Sr$_2$RuO$_4$. 
 
The pairing state symmetry is determined at $T_c$  by the eigenvector 
of Eq.(16) corresponding to 
$\lambda_{\mu} (T)=1$. As well as in the case of superfluid 
$^3$He many other possible states besides the $A$-phase can occur.
Note, that recent
experiments indicate the occurrence of the line nodes in the superconducting
state of Sr$_2$RuO$_4$. These are the power rather an exponential 
laws of the specific heat, $C(T) \propto T^2$ \cite{62}, spin-lattice
relaxation rate $1/T_1 \propto T^3$ \cite{64}, the heat capacity $\kappa
\propto T^2$ \cite{65}, and ultrasound attenuation \cite{67}. Therefore, other
symmetries of the superconducting order parameter like $p$-wave symmetry with
nodes or $f$-wave symmetry which are other 
than simple nodeless $p$-wave symmetry of the superconducting gap have to be
considered. Note, the largest eigenvalue of Eq. (\ref{gaprut}) corresponds 
to a minimum in the free energy and thus  will yield the 
symmetry of the superconducting order parameter $\Delta_l$ in Sr$_2$RuO$_4$.
One could expect that the formation of the node should correspond 
to the influence of the incommensurate antiferromagnetic fluctuations. 
Let us see whether the simple nodeless $p$-wave symmetry is possible if the 
IAF at the wave vector {\bf Q}=$(2\pi/3,2\pi/3)$ are present.

Using appropriate symmetry representations \cite{manfred} we discuss 
the solutions of the gap equation assuming $p$, $d$, or $f$-wave 
symmetry of the order parameter in the RuO$_2$-plane \cite{ilya1}: 
\begin{eqnarray} 
{\bf d}_p({\bf k}) & = & \Delta_0 {\bf \hat{z}} 
(\sin k_x a + i \sin k_y a), \\ 
\Delta_d({\bf k}) & = & \Delta_0 (\cos k_x a -\cos k_y a), \\ 
{\bf d}_{f_{x^2-y^2}}({\bf k}) & = & \Delta_0 {\bf \hat{z}} 
(\cos k_x a -\cos k_y a)(\sin k_x a + i \sin k_y a). 
\label{symmetry} 
\end{eqnarray} 
Here the $\hat{\bf z}$-unit vector refers to the $d_z$ component  
of the Cooper-pairs as observed in the 
experiment \cite{izawa}. We will present a 
possible explanation for this below.  
These symmetries of the superconducting order parameter   
must be substituted into the Eq.(\ref{gaprut}).  
For simplicity we consider first the solution of Eq. (\ref{gaprut})  
in the RuO$_2$-planes and then discuss what is happening along 
the $c$-axis.     
 
Solving Eq. (\ref{gaprut}) in the first BZ down to 5 K we have  
found that $p$-wave symmetry yields the largest energy gain  
for the $xy-$band, while for the $xz-$ and $yz-$bands the situation  
is more complicated.   
We find that the  
gap equations for the $xy$- and $yz-$, $xz$-bands can be separated due 
to their weak interaction. The result is that 
for the $xy$-band the $p$-wave is the most stable  
solution, while for the $xz-$ and $yz-$bands $f_{x^2-y^2}$-wave  
symmetry yields the largest eigenvalue due to stronger nesting. 

\begin{figure} 
\begin{center} 
\includegraphics[width=0.5\textwidth,angle=0]{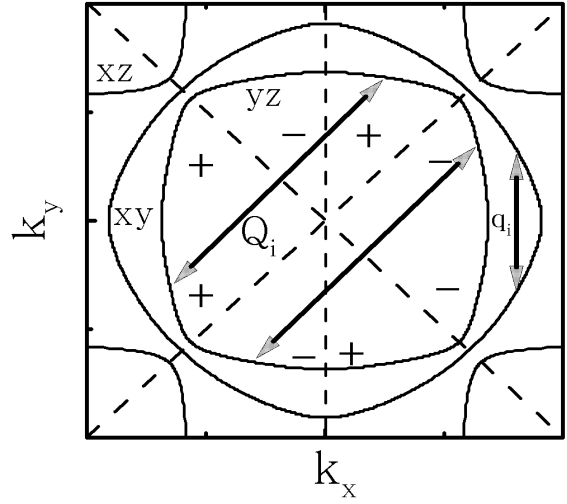} 
\end{center} 
\caption{Symmetry analysis of the order parameter for 
  triplet pairing in the first BZ for $k_z$=0 (solid curves) and 
  $k_z = \pi/2c$ (dotted curves). $\alpha$, $\beta$, and $\gamma$ 
  denote the FS of the corresponding hybridized bands. The wave 
  vectors {\bf Q}$_i$ and {\bf q}$_i$ are the main wave vectors 
  contributing to the susceptibility (without spin-orbit coupling).  
  For $f_{x^2 -y^2}$-wave symmetry the nodes of the real part 
  of the order parameter are shown (dashed lines) and the 
  regions + ( - ) where the $f_{x^2-y^2}$-wave superconducting gap is positive 
  (negative). Note, for the real part of a $p$-wave order parameter   
  the node would occur only along $k_x $=0. 
} 
\label{fermif} 
\end{figure} 
How to understand the competition between $p$- and $f$-wave pairing?. For 
doing this we display in
Fig. \ref{fermif} characterizes the solutions of Eq. (\ref{gaprut}).  
Note, a good approximation is to linearize Eq. 
(\ref{gaprut}) with respect to  
$\Delta^{i}_l$, i.e. $E_{{\bf k}'}^{i} \rightarrow 
\epsilon_{{\bf k}'}^{i}$. We use also $ \tanh (\epsilon_{{\bf 
    k}'}^{i}/2k_B T)=1$. 
Thus, the main contribution to the 
pairing comes from the states at the Fermi level. We present our 
results in terms of  the Fermi surface of the RuO$_2$ plane. The   
wave vectors {\bf Q}$_i$ and  
{\bf q}$_i$ refer to the peaks in $\chi({\bf q}, \omega)$.  
The 
areas with $\Delta_{f_{x^2-y^2}}>0$ and  
$\Delta_{f_{x^2-y^2}}<0$ are denoted by (+) and (-), 
respectively.   
Note, that the minus sign in Eq. (\ref{gaprut}) is absent for 
triplet pairing.  The summation 
over ${\bf k}'$ in the first BZ is dominated by the contributions due 
to ${\bf Q}_i$ for the $\alpha$- and $\beta$-bands  
and the one due to ${\bf q}_i$ for the $\gamma$-band.  
 
As can be seen from Fig. \ref{fermif} in the case of  
$f_{x^2-y^2}$-wave symmetry for the $xy-$band  
the wave vector ${\bf q}_i$ bridges the same 
number of portions of the FS with opposite and equal sign.   
Therefore, for the $xy$-band no solution with $f_{x^2-y^2}$-wave 
symmetry is possible. On the other hand, for the $xz-$ and $yz$-band 
the wave vector  
${\bf Q}_i$ bridges portions of the FS with {\it equal} signs of the 
$f_{x^2-y^2}$-superconducting order parameter. 
The eigenvalue of this order parameter is  
also enhanced due the interband nesting for $xz$ and  
$yz$-bands. Thus, superconductivity in the  
$xz$- and $yz$-band in the RuO$_2$ plane  
is indeed possible with $f_{x^2-y^2}$-wave order parameter.  
Therefore, solving Eq. (\ref{gaprut}) for the three band picture we 
found a competition between triplet $p$-wave and  
$f_{x^2-y^2}$-wave superconductivity in the RuO$_2$-plane. 
 
We can immediately rule out singlet Cooper-pairing.  
In particular, assuming $d_{x^2 -y^2}$-symmetry for Sr$_2$RuO$_4$ 
the 
Eq. (\ref{gaprut}) yields an eigenvalue which is lower than in  
the case of triplet pairing.  
As can be seen using Fig. \ref{fermif} this is plausible.  
Note, we get for d$_{x^2-y^2}$-wave symmetry a change of sign of  
the order parameter upon crossing the diagonals of the BZ.   
According to Eq. (\ref{gaprut}) wave vectors around 
{\bf Q}$_i$ connecting areas (+) and (-) contribute constructively to 
the pairing. Contributions due to {\bf q}$_i$ and the background 
connecting areas with the same sign subtract from the pairing  
(see Fig. \ref{fermif} with nodes at the diagonals).  
Therefore, we find that the four contributions  
due to {\bf q}$_i$ in the $xy-band$ do not allow 
$d_{x^2-y^2}$-wave symmetry in the $xy-$band. Despite the  
pair-building contribution due to {\bf Q}$_i$ one finds that the eigenvalue  
of the $d_{x^2-y^2}$-wave symmetry in the $xy$-band is smaller  
than for the $f_{x^2-y^2}$-wave symmetry. This is due to the  
large contribution from  
{\bf Q}$_i$ to the cross-terms for the triplet pairing which are absent  
for singlet pairing.   
For $d_{xy}$-symmetry where the nodes are along ($\pi$,0) and (0,$\pi$) 
directions we argue similarly. Thus, we exclude this symmetry.   

The $f_{x^2 - y^2}$-wave symmetry solution has vertical
lines of nodes along the $k_z$-axis. In principle, this has to be observable 
in the
strong four-fold anisotropy of the thermal conductivity $\kappa(\theta, H)$ in
applied magnetic field along $ab$-plane. One 
has to admit, however, the recent measurements 
have shown only a weak anisotropy \cite{65} 
that suggests the line nodes in Sr$_2$RuO$_4$ are not vertical.

A phenomenological mechanism to produce horizontal line nodes has been proposed
by Zhitomirskiy and Rice \cite{rice}. As we have already mentioned,
the magnetic fluctuations are different with respect to the various bands.
The ferromagnetic fluctuations are originating from the $xy$-band while 
the incommensurate antiferromagnetic fluctuations are due to the nesting of
$xz$ and $yz$-bands. In this case it was proposed  
that in the active $xy$-band
the superconducting gap has no nodes and can be described by the nodeless 
$p$-wave gap. Then, due to interband scattering of the Cooper-pairs the
superconductivity is induced in the $xz$ and $yz$-bands with the symmetry
\begin{equation}
{\bf d_2 (k)} \sim \left( \sin \frac{k_x a }{2} \cos \frac{k_y a }{2} 
+ \sin \frac{k_y a }{2} \cos \frac{k_x a }{2} \right) \cos \frac{k_z c}{2} 
\quad.
\label{gapc}
\end{equation}
Note, this gap has the horizontal line of nodes at $k_z = \pm \pi/c$.
Then even a combination of the nodeless $p$-wave symmetry and that with nodes 
will still be present, only the position of the nodes will be slightly shifted
along $c$-axis. This behavior correctly describes the temperature dependence 
of the specific heat below $T_c$ \cite{rice}. 
A similar model was proposed by Annett {\it et al.} \cite{litak}. 
In this model there are two distinct attractive pairing interactions. 
One, acting between nearest neighbor in-plane $xy$ orbitals, leads to a 
nodeless chiral $p$-wave state on the $\gamma$-band. The second acts between 
nearest neighbors between planes, and is assumed to act primarily on the 
d$_{xz}$ and d$_{yz}$ orbitals, which have lobes perpendicular to the 
planes. This second interaction produces a gap with line nodes
on the $\alpha$ and $\beta$ Fermi surface sheets. This gap model was 
shown to be in quite good agreement with experiments on specific 
heat \cite{62}, penetration depth \cite{67}
and  thermal conductivity \cite{izawa}.  Similar to the Zhitomirskiy 
Rice approach \cite{rice}, this 
model is also 'orbital dependent superconductivity', but in this case  
there are no 'active' and 'passive' bands, since all bands have a 
pairing interaction and the gap is of similar magnitude on all 
three Fermi surface sheets.

Alternatively, interlayer coupling models have been 
proposed \cite{hase,yana,shigeru}. 
These also give rise to gap functions of the form of Eq. (\ref{gapc}), 
either on one or all of the three Fermi surface sheets. The physical 
mechanisms responsible for the interlayer coupling and hence the gap 
node are assumed to be either interlayer Coulomb interactions, or 
dipole-dipole interactions.

The question remains whether spin fluctuations can also explain the formation
of the nodes along the $c$-axis in the 'passive' $xz$ and $yz$-bands. 
In order to explain this we have first to take into account the magnetic
anisotropy observed by NMR in the normal state \cite{imad}. 
In the following we will show that 
spin-orbit coupling is an important interaction in the physics of Sr$_2$RuO$_4$
and we shall argue that it leads to the formation of the 
horizontal line nodes in the superconducting state.

\subsection{Spin-orbit coupling effects in the normal and 
superconducting states}

So far we investigated the three-band Hubbard Hamiltonian.  
This model neglects spin-orbit coupling and would 
fail to explain the magnetic anisotropy at low temperatures observed
in NMR experiments \cite{imad}.  Recently it has been proposed 
that the spin-orbit
coupling may play an important role in the superconducting state of
Sr$_2$RuO$_4$. In particular, it was shown by Ng and Sigrist \cite{ng2} that the
spin-orbit coupling lowers the free-energy of the chiral superconducting order
parameter. Another indication of the importance of a spin-orbit 
coupling is the
recent observation of the large spin-orbit coupling in the related
insulating compound Ca$_2$RuO$_4$ \cite{saw}.
Therefore, we extend the theory by adding to the Hubbard 
Hamiltonian the spin orbit coupling:
\begin{equation}
H_{s-o} = \lambda \sum_i {\bf l}_i \cdot {\bf s}_i
\quad,
\label{spinorbitsr}
\end{equation}
Here, the effective angular momentum {\bf l}$_i$ operates on the three
$t_{2g}$-orbitals on the site $i$ and ${\bf s}_i$ are the electron
spins.  As in an earlier approach of Ref. \cite{ng2} we restrict ourselves to
the three orbitals, ignoring $e_{2g}$-orbitals and choose the coupling
constant $\lambda$ such that the t$_{2g}$-states behave like an $l=1$
angular momentum representation.  Moreover, it is known that the
quasi-two-dimensional $xy$-band is separated from the
quasi-one-dimensional $xz$- and $yz$-bands. Then, one expects that the
effect of spin-orbit coupling is relatively small for the $xy$-band and can be
neglected in the first approximation \cite{ilya2}. 
Therefore, we consider the effect of the
spin-orbit coupling on $xz$- and $yz$-bands only.  Then, the kinetic
part of the Hamiltonian $H_t +H_{so}$ can be diagonalized and the new
energy dispersions are obtained \cite{ilya2}. Note, we use $\lambda = 80$meV 
in accordance with earlier estimations \cite{ng2,saw}.

Most importantly, 
the spin-orbit coupling does not break the time-reversal symmetry and
therefore the Kramers degeneracy between the spin $up$ and $down$ is
not removed.  The resultant Fermi surface consists of three sheets as
observed in experiment \cite{macki}. Then spin-orbit
coupling together with the Hubbard Hamiltonian leads to a new
quasiparticle which we label with pseudo-spin and pseudo-orbital
indices. Note, that despite the spin-orbit coupling causing the spin and 
orbit
quantum numbers not to be good ones we can still identify the
Cooper-pairing to be triplet or singlet. This refers now then to the
pseudo-spin quantum numbers. At the same time, the magnetic behavior
of Sr$_2$RuO$_4$ becomes very anisotropic due to the fact that both
one-particle Green's functions and Land\'e $g$-factors will be
different if the magnetic field is applied along the $c$-axis or in the
$ab$-RuO$_2$ plane.
The anisotropy
arises mainly from the calculations of the Land\'e $g$-factors and, in
particular, their orbital parts. The factors $g_z = \tilde{l}_z +2s_z$
and $g_+ = \tilde{l}_+ +2s_+$ are calculated using the new
quasiparticle states. The latter consist, for example, of $xz$- and
$yz$-orbitals which in turn are the combinations of the initial
orbital states $|2,+1>$ and $|2,-1>$ mixed due to the crystal field.
Then, the matrix elements $<|l_+|>$, $<|l_-|>$ are zero for the $xz$-
and $yz$-orbitals while the $<|l_z|>$ matrix element is not. Therefore,
the longitudinal components of the spin susceptibility of the $xz$-
and $yz$-band get enhanced in comparison to the transverse one.
One of the interesting questions that we
will analyze later is the effect of spin-orbit coupling on the
antiferromagnetic and ferromagnetic fluctuations. This provides
insights into a microscopic explanation of the pairing mechanism and
allows us to calculate the spatial structure of the superconducting
order parameter.

For the calculation of the transverse, $\chi_l^{+-}$, and
longitudinal, $\chi_l^{zz}$, components of the spin susceptibility of
\begin{figure}
\begin{center}
\includegraphics[width=0.5\textwidth]{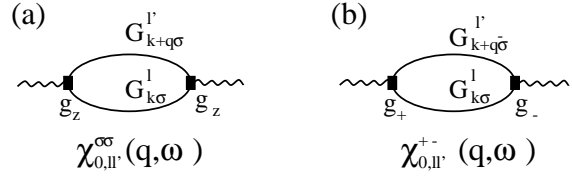}
\end{center}
\caption[]{Diagrammatic representation of (a) the longitudinal 
and (b) the transverse magnetic susceptibility. The full lines
represent the electron Green's function with pseudo-spin $\sigma$
and pseudo-orbital $l$ quantum numbers.  The Lande's $g$-factors are
denoted by $g_{+} = l_+ + 2 s_+$ ($g_-=l_- + 2s_-$) and $g_{z} = l_z
+ 2s_z$.}
\label{chisrdiagram}
\end{figure}
each band $l$ we use the diagrammatic representation shown in Fig.
\ref{chisrdiagram}. Note that because the Kramers degeneracy is 
not removed by 
spin-orbit coupling $G_{k,+-}^l =G_{k,zz}^l$. 
Thus, as we have already mentioned before, 
the anisotropy
arises mainly from the calculations of the Lande's $g$-factors and in
particular their orbital parts. The factors $g_z = \tilde{l}_z +2s_z$
and $g_+ = \tilde{l}_+ +2s_+$ are calculated using the new
quasiparticle states. The latter consist, for example, of $xz$- and
$yz$-orbitals which in turn are the combinations of the initial
orbital states $|2,+1>$ and $|2,-1>$ mixed due to the crystal field.
Then, the same matrix elements as above are zero for the $xz$-
and $yz$-orbitals while $<|l_z|>$ matrix element is again not. Therefore, 
the longitudinal components of the spin susceptibility of the $xz$- 
and $yz$-band get enhanced in comparison to the transverse one.  We 
obtain, for example, for the $|xz\rangle$-states for the transverse 
susceptibility 
\begin{eqnarray} 
\hspace*{-0.5cm}\chi_{0,xz}^{+-} ({\bf q}, \omega) 
& = &  
- \frac{4}{N} \sum_{\bf k} 
(u_{2{\bf k}}u_{2{\bf k+q}}-v_{2{\bf k}} 
v_{2{\bf k+q}})^2 %\nonumber \\ 
%& \times & 
\, \frac{f(\epsilon_{{\bf k}xz}^{+}) 
-f(\epsilon_{{\bf k+q}xz}^{-})} 
{\epsilon_{{\bf k}xz}^{+} - 
\epsilon_{{\bf k+q}xz}^{-} +\omega +iO^+} 
\, , 
\nonumber \\ 
& & 
\label{lindpm} 
\end{eqnarray} 
and for the longitudinal susceptibility 
\begin{eqnarray} 
\chi_{0,xz}^{zz} ({\bf q}, \omega) 
& = & 
\chi_{xz}^{\uparrow}({\bf q}, \omega) +  
\chi_{xz}^{\downarrow}({\bf q}, \omega) \nonumber\\ 
& \hspace*{-1cm}= & 
\hspace*{-0.5cm}- \frac {2}{N} \sum_{\bf k} 
\left[u_{2{\bf k}} u_{2{\bf k+q}} +  
v_{2{\bf k}} v_{2{\bf k+q}} + 
\sqrt{2}( u_{2{\bf k}} v_{2{\bf k+q}}  
+ v_{2{\bf k}}  
u_{2{\bf k+q}})\right]^2 \nonumber \\ 
& \times & 
\frac{f(\epsilon_{{\bf k}xz}^{+})- 
f(\epsilon_{{\bf k+q}xz}^{+})} 
{\epsilon_{{\bf k}xz}^{+} - 
\epsilon_{{\bf k+q}xz}^{+} +\omega +iO^+}  
\quad. 
\label{lindzz} 
\end{eqnarray} 
Here, $f(x)$ is again the Fermi function and $u_{\bf k}^2$ and 
$v_{\bf k}^2$ are the corresponding coherence factors \cite{ilya2}  
 For all other orbitals the calculations are similar and 
straightforward.
 
Then, one gets within RPA the following expressions for the transverse  
susceptibility 
\begin{eqnarray} 
\chi^{+-}_{RPA,l}({\bf q}, \omega) & = &  
\frac{\chi^{+-}_{0,l}({\bf q}, \omega)}{1- 
U\chi^{+-}_{0,l}({\bf q}, \omega)} 
\quad, 
\label{RPApm} 
\end{eqnarray} 
and for the longitudinal susceptibility  
\begin{equation} 
\chi^{zz}_{RPA,l}({\bf q}, \omega) =   
\frac{\chi^{zz}_{0,l}({\bf q}, \omega)}{1-U\chi^{zz}_{0,l}({\bf q}, \omega)} 
\quad, 
\label{RPApzz} 
\end{equation} 
where $\chi^{zz}_{0,l}= \chi^{++}_{0,l} + \chi^{--}_{0,l}$.   
These susceptibilities are used in the corresponding  
pairing interaction for triplet pairing. In order to compare our 
results with NMR and INS experiments we take 
\begin{equation} 
\chi_{tot}^{zz} = \sum_{l}\chi_{RPA,l}^{zz} 
\end{equation} 
and 
\begin{equation} 
\chi_{tot}^{+-} = \sum_{l}\chi_{RPA,l}^{+-} 
\quad . 
\end{equation}

\begin{figure}[h] 
\begin{center} 
\includegraphics[width=0.9\textwidth,angle=0]{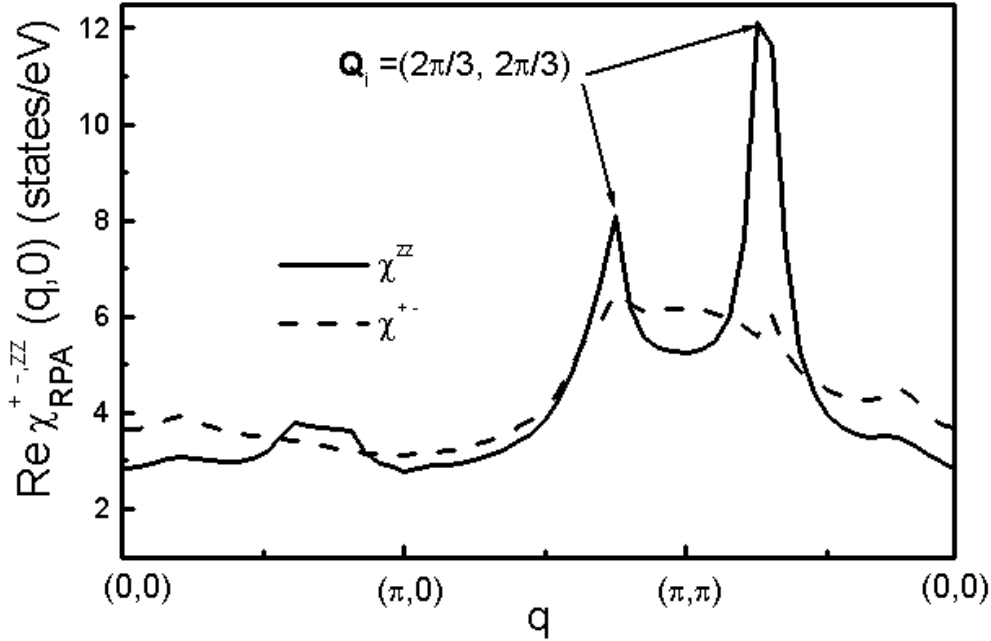} 
\end{center} 
\caption{ Calculated real part of the longitudinal and  
transverse components of the total RPA spin susceptibility  
$\chi^{+-,zz}_{RPA} = \sum_l \chi^{+-,zz}_{l,\, RPA}$ ($l$ refers to 
the band indices) in the normal state of Sr$_2$RuO$_4$ 
along the route $(0,0)\to(\pi,0)\to(\pi,\pi)\to(0,0)$ in 
the first BZ.  Due to spin-orbit interaction the transverse component 
($\chi^{+-}$)  
doe not contain incommensurate antiferromagnetic fluctuations (IAF) 
from $xz$ and $yz$-orbitals, while in the longitudinal 
component ($\chi^{zz}$) they are enhanced. 
} 
\label{ruthimit} 
\end{figure}

Using the random phase approximation(RPA) for each particular band 
we calculate the longitudinal and 
transverse components of the total susceptibility in the 
RuO$_2$-plane. Its real parts at $\omega=0$ are shown in Fig. \ref{ruthimit}. 
As a result of the spin-orbit coupling the magnetic response becomes 
very anisotropic along the whole Brillouin Zone. Since the spin and 
orbital degrees of freedom are mixed now, the orbital anisotropy will 
be reflected in the magnetic susceptibility. As one sees in   
Fig. \ref{ruthimit} the longitudinal component has a 
pronounced peak at {\bf Q}$_i$, while the transverse one does not show 
these features at all and is almost isotropic (it reproduces mostly the 
response of the $xy$-band). In order to understand why 
the longitudinal susceptibility shows mostly the nesting features of 
the $xz$- and $yz$-bands one has to remember that due to spin-orbit 
coupling the orbital component of the magnetic susceptibility cannot 
be neglected. Therefore, the matrix elements such as $<i|l_z|j>$ and  
$<i|l_{+(-)}|j>$ have to be taken into account. At the same time the $xz$- 
and $yz$-bands consist of the $|2,+1>$ and $|2,-1>$ orbital 
states. One sees that while the longitudinal component 
gets an extra term due to $l_z$, the transverse component does 
not, since the transitions between states $|2,+1>$ and $|2,-1>$  are not 
allowed. Thus, the contribution of the nesting of the $xz$ and $yz$ 
orbitals to the longitudinal component of the susceptibility is larger 
than to the transverse one and incommensurate antiferromagnetic  
fluctuations are almost absent.   
\begin{figure}[h] 
\begin{center} 
\includegraphics[width=0.9\textwidth,angle=0]{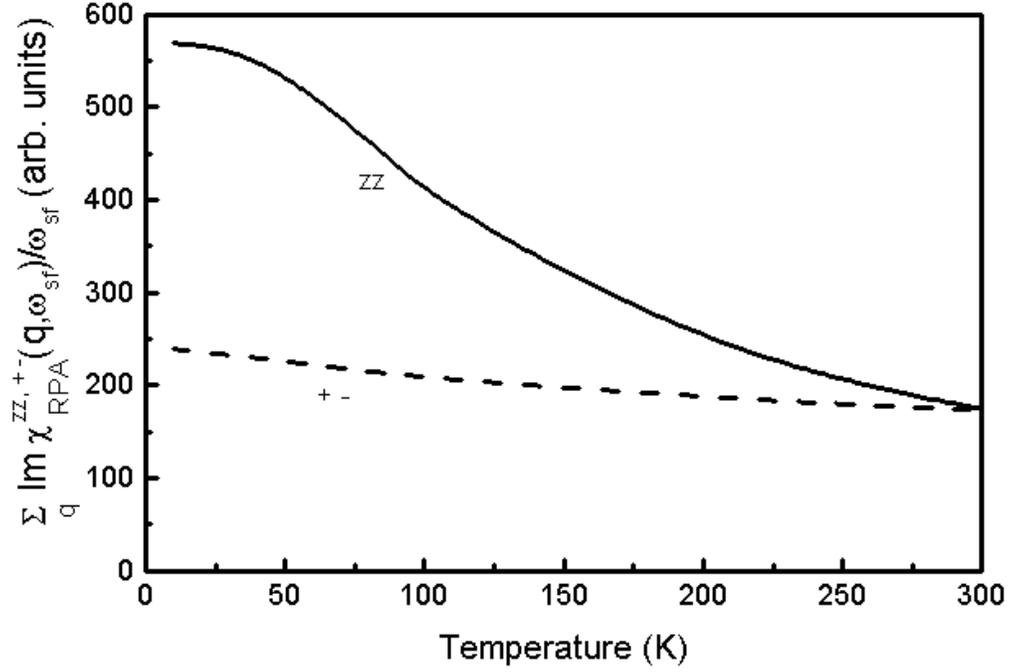} 
\end{center} 
\caption{Temperature dependence of the magnetic anisotropy reflected by 
the imaginary part of the  
spin susceptibility divided by $\omega_{sf}$ and summed over {\bf q}. Note,   
$zz$ and $+-$ refer to the out-of-plane (solid curve)  
and in-plane (dashed curve) 
components of the RPA spin susceptibility.
} 
\label{rutimhi} 
\end{figure} 
\begin{figure}[h] 
\begin{center} 
\includegraphics[width=0.85\textwidth,angle=0]{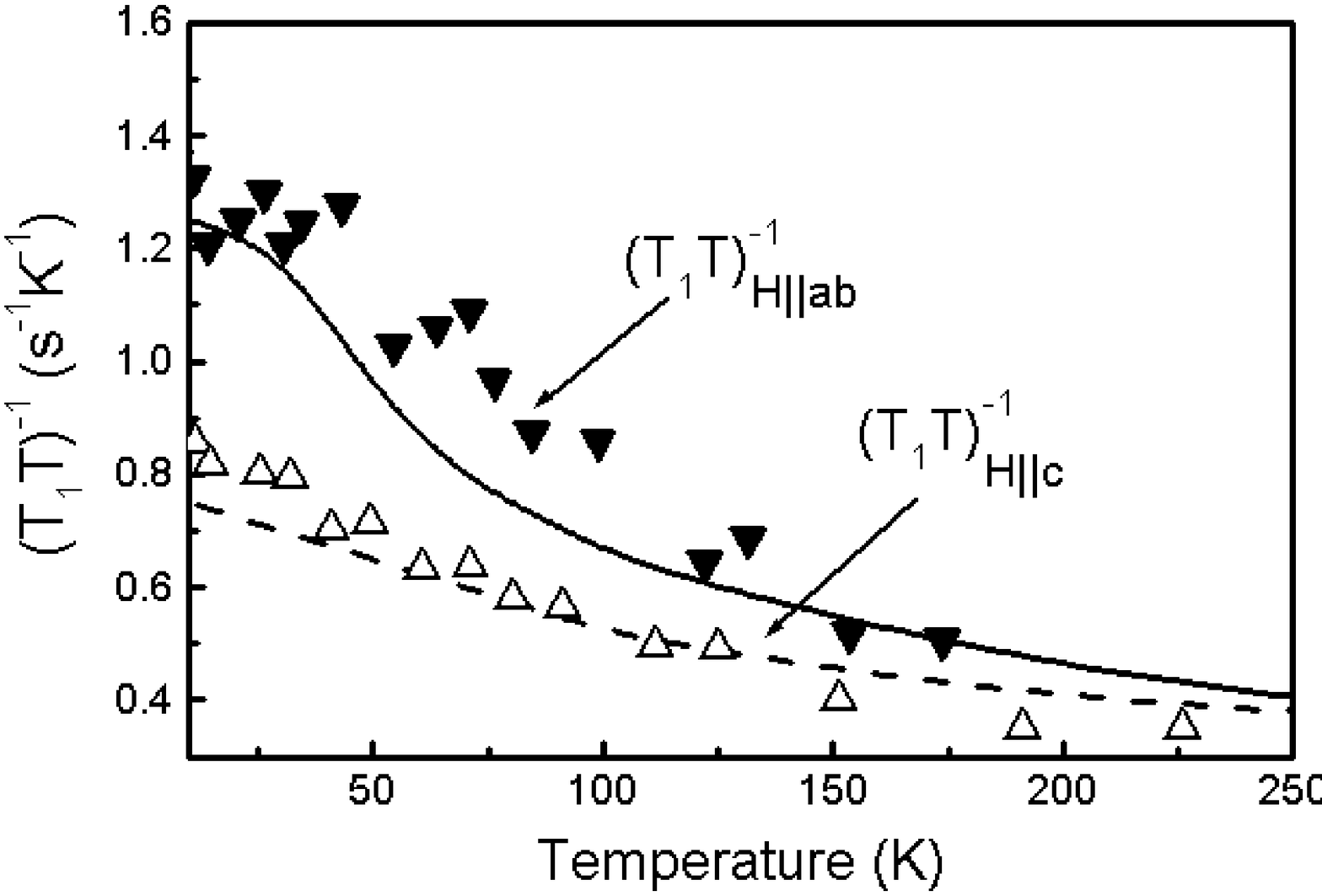} 
\end{center} 
\caption{ Calculated normal state temperature dependence of the  
nuclear spin-lattice relaxation rate $T_1^{-1}$ of $^{17}$O in the  
RuO$_2$-plane for an   
external magnetic field applied along $c$-axis (dashed curve)  
and along the $ab$-plane (solid curve) using Eq. (31). 
Triangles are  
experimental results taken from Ref. \protect\cite{ishida}.  
} 
\label{t1tris} 
\end{figure} 

In order to demonstrate the temperature dependence of the magnetic  
anisotropy induced by the spin-orbit coupling we display in 
Fig. \ref{rutimhi}  the  
temperature dependence of the quantity 
\begin{displaymath}   
\sum_{\bf q} \frac{Im \chi_{RPA} ({\bf q}, \omega_{sf})}{\omega_{sf}} 
\end{displaymath} 
for both components. At room temperature both longitudinal and transverse  
susceptibilities are almost identical, since thermal effects wash out the  
influence of the spin-orbit interaction. With decreasing temperature  
the magnetic anisotropy increases and at low temperatures we find  
the important result that the out-of-plane  
component $\chi^{zz}$ is about two times larger than the in-plane one:  
\begin{displaymath} 
\chi^{zz}>\chi^{+-}/2 \quad. 
\end{displaymath} 

We also note that our results are in accordance with earlier estimations  
made by Ng and Sigrist \cite{ng1}. However, there is  
one important difference. In their work  
it was found that the IAF are slightly enhanced in the longitudinal  
components of the $xz$- and $yz$-bands in comparison to the transverse one. 
In our case we have found that the longitudinal  
component of the magnetic susceptibility is strongly enhanced due to  
orbital contributions. Moreover, we have shown by taking into account  
the correlation  
effects within RPA that the IAF are further  
enhanced in the $z$-direction. 

Finally,  
in order to compare our results with experimental data we calculate the  
nuclear spin-lattice relaxation rate $T_1^{-1}$ for the  
$^{17}$O ion in the RuO$_2$-plane for different external  
magnetic field orientation $(\mbox{i = }a,b, \mbox{ and   }c)$:  
\begin{eqnarray} 
\left[\frac{1}{T_1 T}\right]_i & = & \frac{2k_B \gamma^2_n}{(\gamma_e \hbar)^2} 
\sum_{\bf q} |A_{\bf q}^{p}|^2 \frac{\chi''_{p} ({\bf q}, \omega_{sf})} 
{\omega_{sf}} 
\quad, 
\label{t1t} 
\end{eqnarray}   
where $A_{\bf q}^{p}$ is the  
$q$-dependent hyperfine-coupling constant perpendicular to the  
$i$-direction.  
 
In Fig. \ref{t1tris} we show the calculated temperature dependence  
of the spin-lattice relaxation for an external magnetic field  
within and perpendicular to the RuO$_2$-plane together with experimental data.  
At $T = 250$ K the spin-lattice relaxation rate is almost isotropic.  
Due to the anisotropy in the  
spin susceptibilities arising from spin-orbit coupling  
the relaxation rates become increasingly  
different for decreasing temperature. The largest anisotropy occurs  
close to the superconducting transition temperature in  
good agreement for experimental data \cite{ishida}. 

Note, most recently similar magnetic anisotropy has been observed by means of 
inelastic neutron scattering \cite{brade}. However, the absolute 
magnitude of the anisotropy seems to be smaller than in the case of the 
NMR experiments \cite{ishida}. The reason for this might be that in the NMR
experiments the magnetic susceptibility is summed over the whole Brillouin
Zone while in the INS only its part around ${\bf Q}=(2\pi/3,2\pi/3)$ is 
probed.

What is the influence of spin-orbit coupling for determining the symmetry of
the superconducting gap in Sr$_2$RuO$_4$?
\begin{figure} 
\begin{center} 
\includegraphics[width=0.7\textwidth,angle=0]{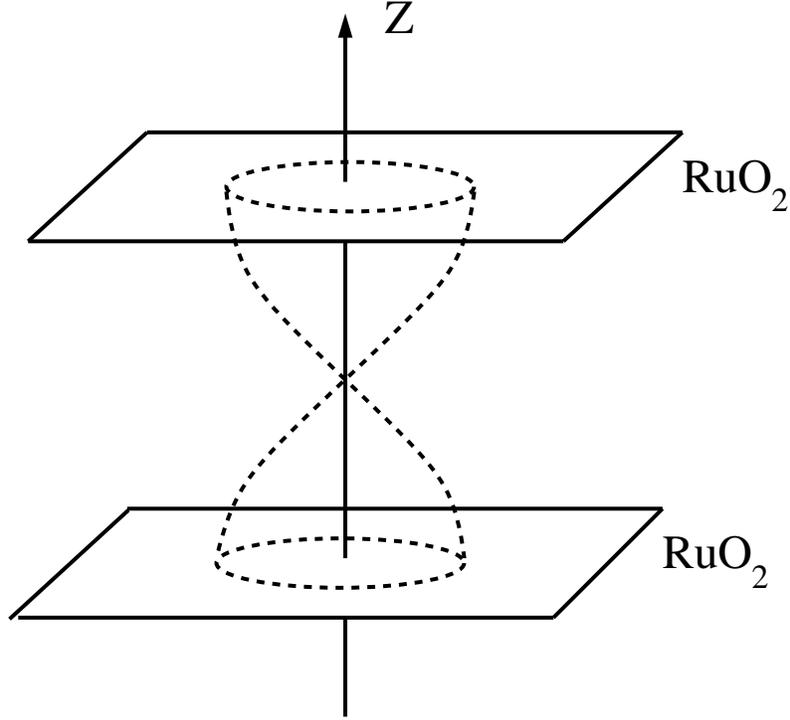} 
\end{center} 
\caption{Schematic representation of the possible node formation in  
  the order parameter  
  between the RuO$_2$-planes (in real coordinate representation)  
  as resulting from the magnetic anisotropy.   
  Here, the amplitude of the order parameter along the $z$-direction   
  has been drawn in  
  cylindrical coordinates between RuO$_2$-planes. This horizontal line node 
 in the superconducting gap resulting from antiferromagnetic spin 
 fluctuations along $z$-direction seems to be in agreement 
  with thermal conductivity measurements below $T_c$ 
  \protect\cite{65}.  
} 
\label{spatial} 
\end{figure} 
First, the spin-orbit 
coupling affects the spin dynamics and as we have shown induces the  
anisotropy in the spin subspace. In particular the two-dimensional 
IAF at {\bf Q}$_i = (2\pi/3,2\pi/3)$ have a polarization  along the  
$z$-direction. This simply means that the antiferromagnetic moments  
associated with these fluctuations align along the $z$-direction. At 
the same time the ferromagnetic fluctuations are in the $ab$-plane. 
This is illustrated in Fig. \ref{spatial}. Since 
the interaction in the RuO$_2$-plane ($k_z$=0) is mainly ferromagnetic  
nodeless $p$-wave pairing is indeed possible and the superconducting 
order parameter in all three bands will not have nodes. This differs 
from previous results in which only the $xy$-band was nodeless. 
Therefore, we safely conclude that triplet $p$-wave pairing 
without line-nodes will be realized in the RuO$_2$-plane.   
On the other 
hand the state between the RuO$_2$-plane will be determined mainly 
by the IAF that are polarized along the $c$-direction. This implies that 
the magnetic interaction between the planes has to be rather 
antiferromagnetic than ferromagnetic.  
This antiferromagnetic interaction strongly suggests  
a line of nodes for $\Delta({\bf k})$ between the 
neighboring RuO$_2$-planes. This is in agreement with the arguments 
presented in Ref. \cite{rice}.
Then, due to the magnetic anisotropy  
induced by spin-orbit coupling a nodeless $p-$wave pairing is possible  
in the RuO$_2$-plane as experimentally observed \cite{65}, while a  
node would lie 
between the RuO$_2$-planes. 

\begin{figure} 
\begin{center} 
\includegraphics[width=0.9\textwidth,angle=0]{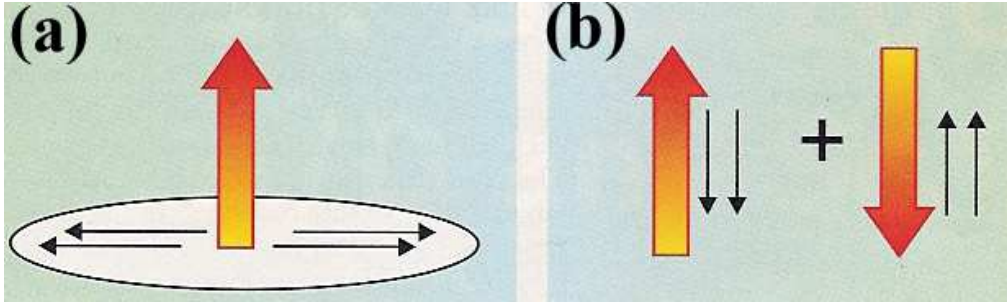} 
\end{center} 
\caption{ Illustration of the possible superconducting states in 
  Sr$_2$RuO$_4$ taken from Ref. \protect\cite{physto}.  
  (a) $\Delta ({\bf k}) =  d_z (\sin k_x + i 
  \sin k_y)$ and (b) $\Delta ({\bf k}) = d_x \sin k_x +  
  d_y \sin k_y$ have different spin and total angular 
  momentum. The  $\Delta ({\bf k}) = d_z (\sin k_x + i 
  \sin k_y)$ superconducting state has an angular momentum along 
  $c$-axis (thick arrow) and the spin is lying in the RuO$_2$-plane. 
} 
\label{triplet11} 
\end{figure}

Note, the spin-orbit coupling determines also the orientation of the 
${\bf d}$-vector of the Cooper-pairs.
In a triplet superconductor, the pairing state can be represented by the 
three-dimensional vector {\bf d(k)}, whose magnitude and direction may 
vary over the Fermi surface in {\bf k}-space. Like in the case of superfluid 
$^3$He, several 
pairing states related to the different orientation of the total spin 
moment of the Cooper-pair $d_z, \, d_y, \, d_z$ 
may have the same condensation energy under weak-coupling conditions.  
This degeneracy is related to the spin rotation symmetry being present 
in $^3$He (which leads to presence of the so-called B-phase),  
but is lifted by the spin-orbit coupling in Sr$_2$RuO$_4$ \cite{ng2}. Then, the
analogy of the $A$-phase-like can be realized in Sr$_2$RuO$_4$ ({\bf d(k)}$_A$ 
  = (0,0,$ d_z(k_x \pm i k_y)$)) while the two-dimensional analog of the 
$B$-phase will be suppressed  
({\bf d(k)}$_B$  = $|d|$ ($ k_x, k_y,$ 0)). 
The different orientation of the spin and orbital angular momenta   
of the Cooper-pair is illustrated in Fig. \ref{triplet11} .  
The important consequence of this order parameter is the uncompensated  
orbital angular momentum of the Cooper-pair (so-called 'chiral' state).  
 
In analogy to $^3$He, interesting effects can be observed by applying 
external magnetic field  or external pressure along $ab$-plane to 
Sr$_2$RuO$_4$. For example, in $^3$He  the transition 
from the $B$ to the $A$ phase is observed under pressure 
while in an external magnetic field the $A$ 
transition splits into $A_1$, $A_2$ phases.  
In Sr$_2$RuO$_4$ the situation is somewhat opposite. Superconductivity in 
Sr$_2$RuO$_4$  is already 
in the $A$-like phase and therefore, it would be interesting to see whether 
by applying pressure or magnetic 
field one could cause a transition into a $B$-like phase. Note that recently 
a second phase has been observed in high magnetic field applied along 
$a$-direction ($H || (100)$) with 
$T_c \sim 0.5 \,\, T_{c0}$ \cite{maenosec}. The origin of this feature is 
not clear at the moment.

\subsection{ Comparison of superconductivity in cuprates and
Sr$_2$RuO$_4$ }

Despite of the similar crystal structures the differences in the electronic 
states close to the Fermi level leads to the different electronic structure of
cuprates and ruthenates. As a result the spin fluctuations are very
different. In both systems the spin fluctuations are strong but while in
cuprates the antiferromagnetic fluctuations at {\bf Q}=$(\pi,\pi)$ in
Sr$_2$RuO$_4$ there is a competition between ferromagnetic and incommensurate
antiferromagnetic fluctuations at {\bf Q}$_{IAF}=(2\pi/3,2\pi/3)$. 
As we have shown earlier the superconductivity in the Sr$_2$RuO$_4$ seems to
be induced by mainly the ferromagnetic fluctuations and the role of IAF is in
the formation of the line nodes in the passive $xz$ and $yz$-bands between the
RuO$_2$-planes. 

\begin{figure}
\begin{center}
\includegraphics[width=0.5\textwidth]{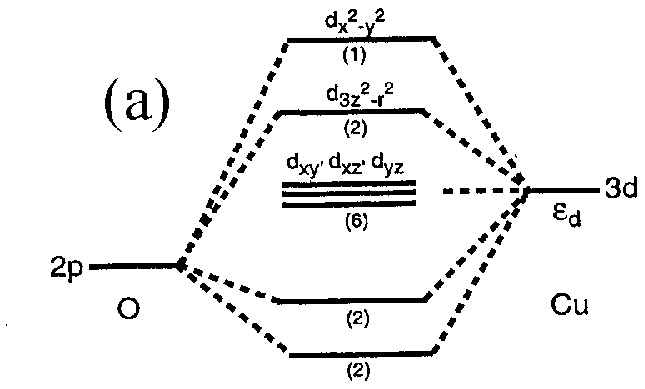}
\includegraphics[width=0.5\textwidth]{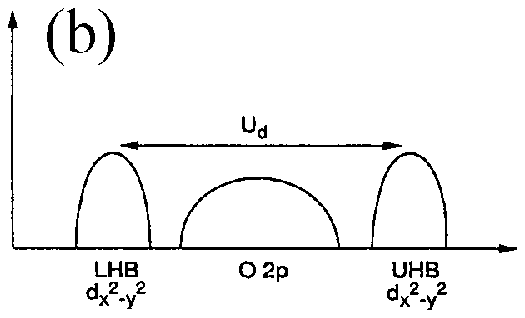}
\end{center}
\caption[]{ Schematic electronic structure of the CuO$_2$-planes of cuprates 
reflecting
bonding between a Cu$^{2+}$ and two O$^{2-}$ ions. (a) Initial
splitting of the bands taking into account crystal field interaction
only. We consider only $d$-electrons (holes) of Cu and $p_x$ and
$p_y$ orbitals of the oxygen. The numbers indicate the occupations
of the different levels in the undoped compound. (b) Role of the
on-site Coulomb repulsion and the splitting of
3d$_{x^2-y^2}$-orbital into a lower (LHB) and upper Hubbard band
(UHB). U$_d$ is the Coulomb repulsion between electrons at Copper.
Due to $\delta=\epsilon_p - \epsilon_d < U_d$ the cuprates are
Mott-Hubbard insulators without doping.}
\label{cuintro}
\end{figure}

The comparison of the superconductivity in ruthenates and cuprates 
was done in Ref. \cite{ovchin} on the
basis of the $t-J-I$ model. For all cuprate compounds the
degeneracy between the copper 
3$d$-orbitals is removed by the lattice structure.
After some straightforward calculations it can be shown 
that the hybridized copper and oxygen orbitals
separate. As shown in Fig. \ref{cuintro}(a), 
the state with highest (lowest) energy has mainly
d$_{x^2-y^2}$-wave character in the undoped cuprates and the $d^9$ copper 
has one missing electron ({\it i.e.} a hole) and 
gives the Cu-ion a spin $\frac{1}{2}$. Thus, in the absence of doping,
the cuprate material is well described by a model of mostly localized
spin-$\frac{1}{2}$ states. The other orbitals are occupied and
therefore can be neglected. However, the Coulomb repulsion
between holes in the same orbital is strong and
must be included. Due to the large Coulomb repulsion the 
d$_{x^2-y^2}$-orbitals split into two so-called lower and upper
Hubbard bands (LHB and UHB), respectively as shown in Fig.
\ref{cuintro}(b). Then, at half-filling the system becomes an
insulator. Furthermore, due to a very large $U_d$ the splitting of
d$_{x^2-y^2}$-band is so large that the oxygen $p$-band lies between
the UHB and the LHB.  The charge transfer gap in cuprates $\delta$
($\delta = \epsilon_p -\epsilon_d$) is smaller than $U_d$, and so underdoped 
cuprates are charge-transfer insulators.
When the cuprates are doped the $t-J$ model appears as an 
effective low-energy Hamiltonian
describing the moving of a hole on the antiferromagnetic
background \cite{zhang}. 

In contrast to the cuprates, in ruthenates the Fermi level lies in the 
center of the $d_{xy}, d_{xz}$, and $d_{yz}$ manifold of states.
In the ruthenates considering the active $xy$-band one
could also consider the moving of the electrons on the ferromagnetic
background. The instability of the system can be analyzed with respect to the
triplet $p$-wave and singlet $d$-wave Cooper-pairing. In a weak-coupling BCS 
theory it was
found \cite{ovchin} that for $p$-wave pairing the van Hove singularity
does not contribute to the $T_c$ since in the effective density of states the
corresponding electron velocity at the Fermi level is cancelled by the angular
dependence of the $p$-wave superconducting gap $\sim(\sin k_x \pm i \sin
k_y)$. At the same time  this
cancellation does not occur for $d_{x^2-y^2}$-wave Cooper pairing 
and thus the van Hove singularity enhances $T_c$.

We can also contrast the spin fluctuation pairing mechanism in the cuprates 
and ruthenates. 
In the strong-coupling Eliashberg-like theory of superconductivity in the
cuprates \cite{khb} the enhancement of $T_c$ occurs
due to pronounced antiferromagnetic spin fluctuations and due to a large 
spectral weight of the pairing interaction at high frequencies. Therefore, by
comparison of superconductivity in Sr$_2$RuO$_4$ and high-$T_c$ cuprates one
may conclude that the higher $T_c$ in cuprates can be explained by the
stronger antiferromagnetic fluctuations and the density of states
effects (van Hove singularity). 
This, however, are only quantitative arguments since a complete
theory of superconductivity in cuprates is still lacking. 

\section{\bf Mott-insulator transition in Ca$_{2-x}$Sr$_x$RuO$_4$}

Originally, the investigations of ruthenates were concentrated on Sr$_2$RuO$_4$
and SrRuO$_3$, since it was believed that in contrast to cuprates in ruthenates
only ferromagnetic fluctuations take place. However, it was soon realized
that the ruthenates reveal much more interesting behavior. 
In particular, the 
richness of the electronic, magnetic and structural phase transitions in
Ca$_{2-x}$Sr$_x$RuO$_4$ has received a lot of attention recently.

\begin{figure}[h] 
\begin{center} 
\includegraphics[width=0.5\textwidth,angle=0]{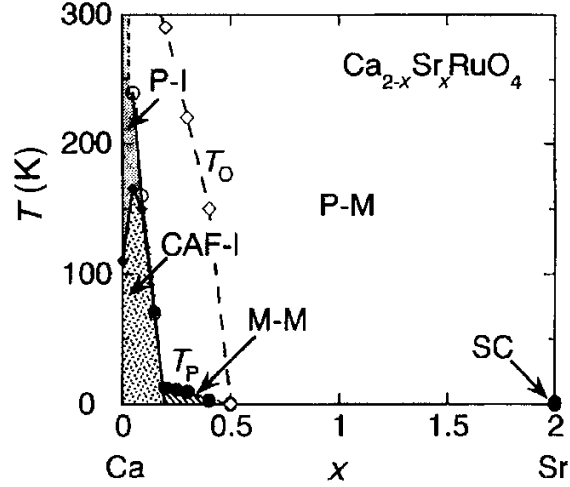} 
\end{center} 
\caption{ Phase diagram of Ca$_{2-x}$Sr$_x$RuO$_4$. P refers to  
paramagnetic, CAF to canted antiferromagnetic, M to magnetic,  
SC to the superconducting phase, -M to the metallic phase, and -I 
to the insulating phase \protect\cite{naka}.  
} 
\label{rutendoping} 
\end{figure}

In particular, in Ca$_{2-x}$Sr$_x$RuO$_4$ a transition from triplet
superconductivity in Sr$_2$RuO$_4$ to an 
antiferromagnetic Mott-Hubbard insulator
in Ca$_2$RuO$_4$ has been found \cite{naka}. For various values of $x$ the
different phase transitions occur as shown in Fig. \ref{rutendoping}. For
example for $x<0.2$ this material is an insulator and for the interval $0.2<x<0.5$ 
Ca$_{2-x}$Sr$_x$RuO$_4$ is a metal with short-range antiferromagnetic
order. At $x \sim 0.5$ there is a crossover which is accompanied by the sharp
enhancement of the ferromagnetic fluctuations in the uniform spin
susceptibility. For $x \to 2$ the system becomes superconducting with
triplet Cooper-pairing.

\begin{figure} 
\begin{center} 
\includegraphics[width=0.5\textwidth,angle=0]{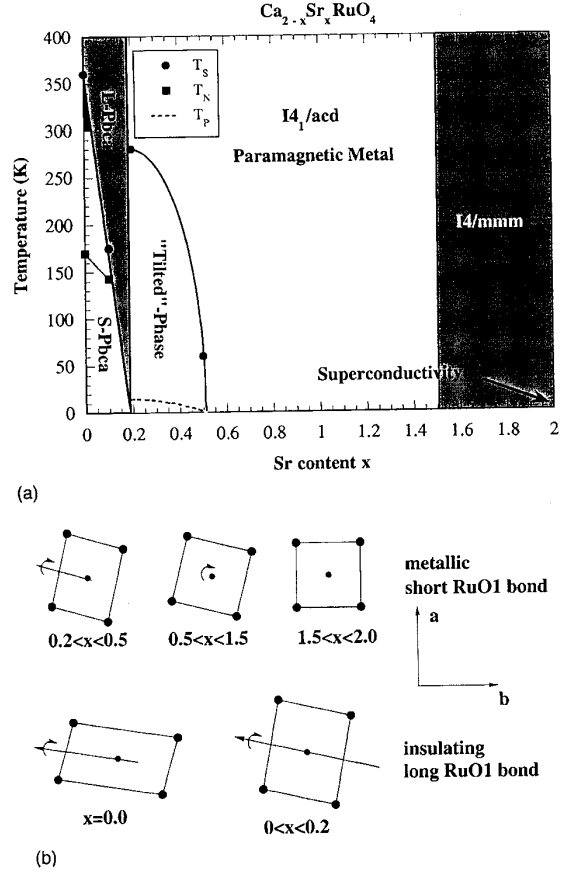} 
\end{center} 
\caption{ Phase diagram of  Ca$_{2-x}$Sr$_x$RuO$_4$ as measured by means of
neutron diffraction \protect\cite{friedt} including the different structural
and magnetic phases and the occurrence of the maxima in the magnetic spin
susceptibility. In the lower part the tilting and rotation distortion of the
octahedra are shown  together with elongation of the basal planes (Ru$-$small
points, O(1)$-$larger points). All phases are metallic except S$-$Pbca. 
} 
\label{friedtfig} 
\end{figure}

A neutron diffraction study on  Ca$_{2-x}$Sr$_x$RuO$_4$ has been made
recently \cite{friedt}, and results in the  
structural phase diagram of this system, shown in Fig. \ref{friedtfig}(a). 
Here, one can see also the experimental data of the structural
($T_S$), magnetic ($T_N$) phase transition and the dashed curve refers to the
temperature $T_p$ where the magnetic susceptibility shows a maximum. 
As one sees for
$x<0.2$ the metal-insulator transition is accompanied by the structural
transition as well where both low and high temperature phases have the
symmetry $Pbca$ with only difference of the parameter $c$ (S-short $c$, L-long
$c$). In the metallic phase there are two structural phase transitions which
are influencing the temperature and concentration dependencies of the magnetic
susceptibility. In Fig. \ref{friedtfig}(b) we illustrate the evolution of the 
structural
transitions with increasing x, starting 
from Sr$_2$RuO$_4$. The rotation of the RuO$_6$
octahedra around $c-$axis leads to a phase $I4_1/acd$. For $x=1$ we have
$\phi= 10.8$. Below $x=0.5$ there are addition rotations around the $ab$-plane
axis which increase the $\theta$ angle and suppression of the octahedra along
the $c$-axis. 

\begin{figure} 
\begin{center} 
\includegraphics[width=0.4\textwidth,angle=0]{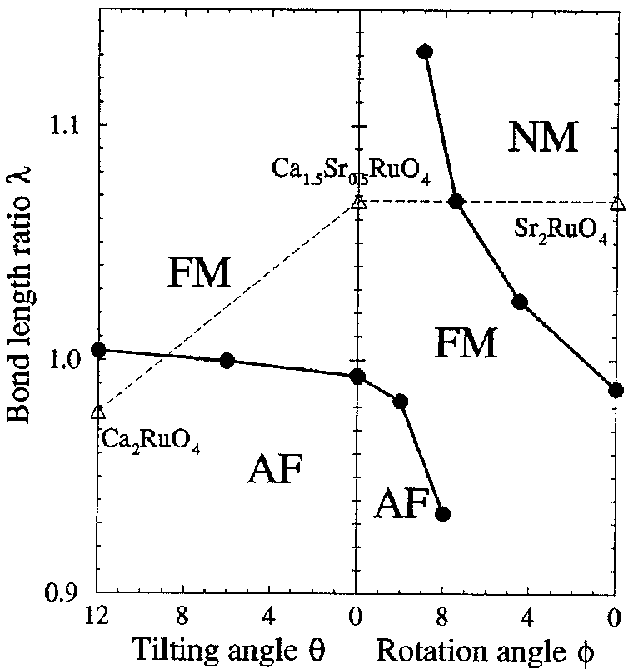} 
\end{center} 
\caption{ The calculated magnetic phase diagram of Sr$_2$RuO$_4$ 
  including structural distortions, taken from 
  Ref. \protect\cite{feng}. The solid lines are calculated phase 
  boundaries, while the triangles linked by dashed lines correspond 
  to the experimental data.  
} 
\label{terakura} 
\end{figure}

An attempt to explain the magnetic phase diagram of Sr$_{2-x}$Ca$_x$RuO$_4$
has been made on a basis of 'ab-initio' band structure
calculations \cite{feng}, as shown in Fig. \ref{terakura}. 
It was shown that the rotations of the RuO$_6$
octahedra around the $c-$axis stabilize the ferromagnetic state, since they
are sufficient to reduce the $pd-\pi$ hybridization between the
$xy$-orbital and the oxygen $2p$-states. 
Then the $xy$-band becomes narrower and
the van Hove singularity shifts towards the Fermi level. At the same time the 
$xz$- and $yz$-bands are only slightly affected. On the contrary, the 
rotation around the 
$ab$-plane changes all three bands completely and that increases
the nesting of $xz$- and $yz$-bands and may stabilize the antiferromagnetic
phase \cite{feng}. 

\begin{figure} 
\begin{center} 
\includegraphics[width=0.5\textwidth,angle=0]{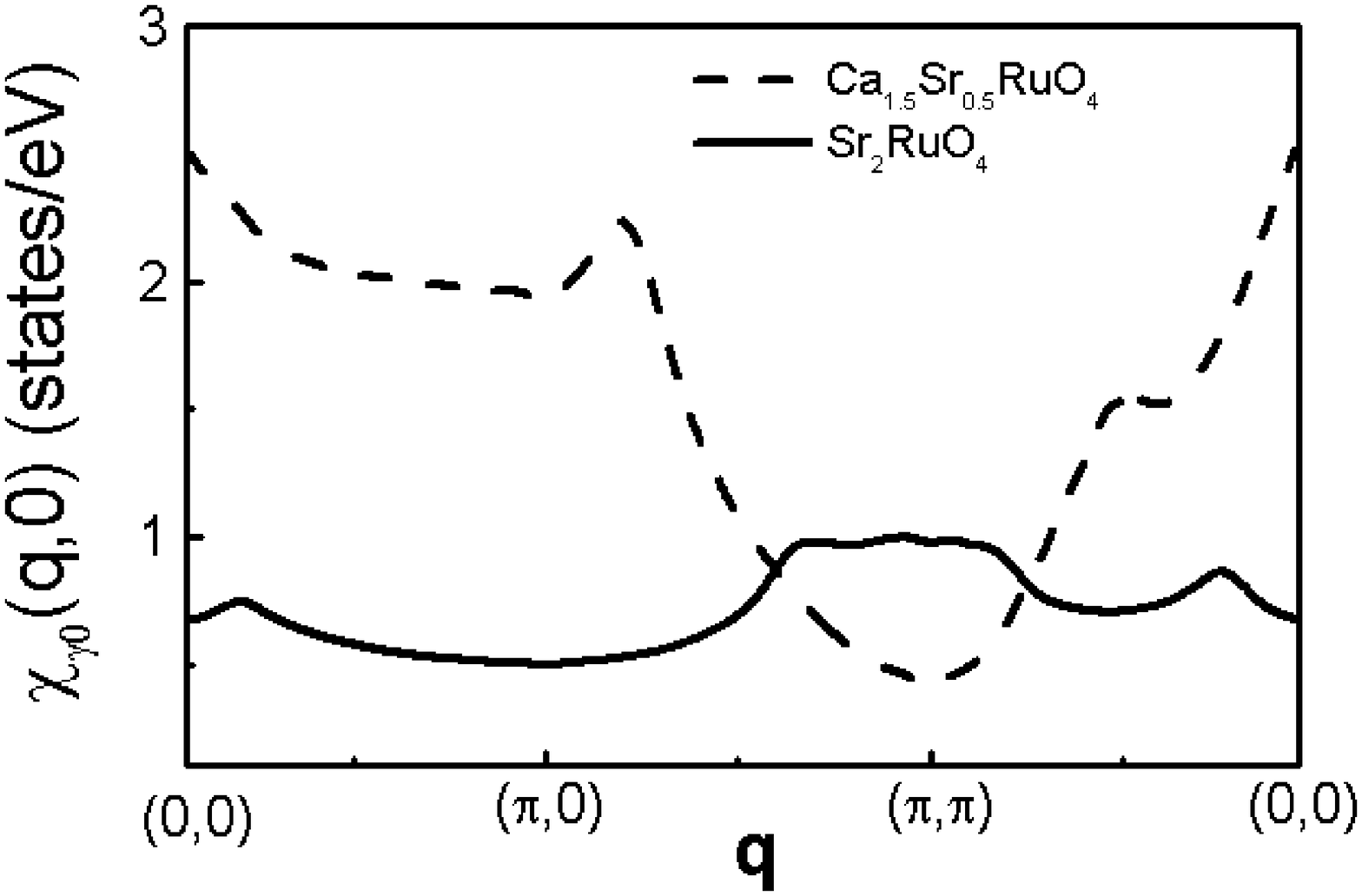} 
\end{center} 
\caption{ The calculated real part of the Lindhard response function,
$\chi_{\gamma 0} ({\bf q},0)$ along the symmetry points in the BZ 
in comparison for the $\gamma$-band for
 Ca$_{0.5}$Sr$_{1.5}$RuO$_4$ and Sr$_2$RuO$_4$. Due to vicinity to the van
Hove singularity the response of the $\gamma$-band is strongly enhanced and
ferromagnetic (around ${\bf q}=0$). 
This is in good agreement with experiment \protect\cite{friedt}.
} 
\label{chiCa} 
\end{figure}

In this picture 
it is easy to understand why the ferromagnetic fluctuation are strongly
enhanced around the so-called critical doping $x \sim 0.5$. Using the LDA
parameters for the tight-binding energy dispersion we calculated the Lindhard
response function of the $xy$-band for Ca$_{0.5}$Sr$_{1.5}$RuO$_4$ and
Sr$_2$RuO$_4$ as shown in Fig. \ref{chiCa}. One could clearly see that due to
vicinity to the van Hove singularity the ferromagnetic response of the 
$xy (\gamma)$-band is strongly enhanced, at ${\bf q}=0$. Furthermore, since the
$\alpha$ and $\beta$-bands are almost unchanged the ferromagnetic response
gets much stronger than the antiferromagnetic ones indicating the dominance of
the ferromagnetic fluctuations around this critical point $x \sim 0.5$.
This is in good agreement with experiment \cite{friedt}.
However, this model cannot explain the Mott-Hubbard transition in 
Ca$_2$RuO$_4$ at finite temperatures, since Ca$_2$RuO$_4$ remains an insulator
even above $T_N$. Therefore, strong electronic correlations has to be taken
into account, as was proposed recently \cite{ovchin1}. Here it was suggested 
that while the 
$xz$- and $yz$-bands are split into the lower and upper Hubbard bands,
the $xy$-band is not split even though 
it is close to the Mott-Hubbard transition. 
The splitting between $xz$($\alpha$)- 
and $yz$($\beta$)-bands
is increased due to interband Coulomb repulsion. 
Indeed, in a mean-field approximation
\begin{equation}
U_{\alpha \beta} n_{\alpha} n_{\beta} \to U_{\alpha \beta} < n_{\alpha} > 
n_{\beta} + U_{\alpha \beta} n_{\alpha} < n_{\beta} >
\label{bandcul}
\end{equation}
the energy dispersions are renormalized
\begin{eqnarray}
\epsilon_{\alpha} \to \, \tilde{\epsilon}_{\alpha} &  = & \epsilon_{\alpha} + 
U_{\alpha \beta} < n_{\beta} > \nonumber \\
\epsilon_{\beta} \to \, \tilde{\epsilon}_{\beta} &  = & \epsilon_{\beta} + 
U_{\alpha \beta} < n_{\alpha} >
\label{bandcul1}
\end{eqnarray}
and the splitting of the $\alpha$- and $\beta$-bands is given by
\begin{eqnarray}
\Delta \epsilon = \tilde{\epsilon}_{\beta} - \tilde{\epsilon}_{\alpha} =
\epsilon_{\beta} - \epsilon_{\alpha} + 
U_{\alpha \beta} (< n_{\alpha} > - < n_{\beta} >) \sim t_{\alpha \beta} +
\frac{2}{3} U_{\alpha \beta} .
\end{eqnarray}
Here we use the data for the filling of the $\alpha$ and
$\beta$-bands from Ref. \cite{singh}. At $U_{\alpha \beta} \approx 1$ eV 
the splitting is 
$\Delta \epsilon \sim 0.8$ eV which is comparable with the band width of these
bands. Note in this case one will have still three Fermi surfaces (as in
LDA calculations) 
due to the $xy$-band, upper Hubbard band of $\beta$-band and one
hole-like Fermi surface from the $\alpha$-band with filling factor
$n_{\gamma}^h \approx 0.3$. 

Due to the rotation of RuO$_6$-octahedra around the $c$-axis the width of the
$xy \, (\gamma)$-band is reduced and the system approaches the critical value
$U \sim U_c$. As we mentioned for the rotation of the RuO$_6$-octahedra around
$c$-axis, the  
$\alpha$ and $\beta$-bands are not changing in this doping range 
($0.5<x<2.0$). Further decreasing $x$ and the rotation of the 
RuO$_6$-octahedra with respect to the $ab$-plane reduces the bandwidth of all
three bands so for $x=0$ the $\gamma$-band is split into the filled 
lower and empty upper Hubbard bands. A similar situation occurs for the
$\beta$-band while the $\alpha$-band is now 
just completely filled. This allows us to
explain the formation of the Mott-Hubbard insulator for Ca$_2$RuO$_4$ in a
way consistent with experimental data of X-ray Absorption
Spectroscopy \cite{mizo}. Note that while LDA+DMFT results \cite{anisim} also 
predict correctly the formation of Mott-Hubbard antiferromagnetic insulator in
Ca$_2$RuO$_4$ the resulting filling of the bands is inconsistent with the
observed data \cite{mizo}. 
 
In addition, a new phase transition in Ca$_2$RuO$_4$ has been found recently 
by applying an external hydrostatic pressure \cite{naka1}. Above a very small 
pressure ($p_c = 0.5$ GPa) Ca$_2$RuO$_4$ shows metallic properties, and 
most unusually it becomes a ferromagnetic metal with a Curie temperature 
$T_c$ strongly dependent on pressure. For example, just above the  
critical pressure $T_c = 12$ K and it reaches its maximum value of $T_c =25$ 
K at 5 GPa. 
In general, in the Hubbard model one would expect a transition from 
the antiferromagnetic insulator to the paramagnetic metal. However, if 
one keeps in mind the multi-band nature of the metallic state, a 
ferromagnetic ordering is possible. Of course, a study of the mechanism 
of magnetic order in the multiband metallic state in Ca$_2$RuO$_4$ should be 
done in the future.

\section{Magnetic versus non-magnetic impurities in Sr$_2$RuO$_4$}

It is known that unconventional superconductors like heavy-fermion compounds
or high-$T_c$ cuprates reveal peculiar behavior if one add magnetic
or non-magnetic impurities. In contrast to conventional $s$-wave
superconductors the non-magnetic as well as the 
magnetic impurities act as strong
pair breakers and suppress the transition temperature $T_c$ of unconventional
superconductors. The suppression of $T_c$ reflects the sensitivity to
translational symmetry breaking and is characteristic of anisotropic
\begin{figure} 
\begin{center} 
\includegraphics[width=0.65\textwidth,angle=0]{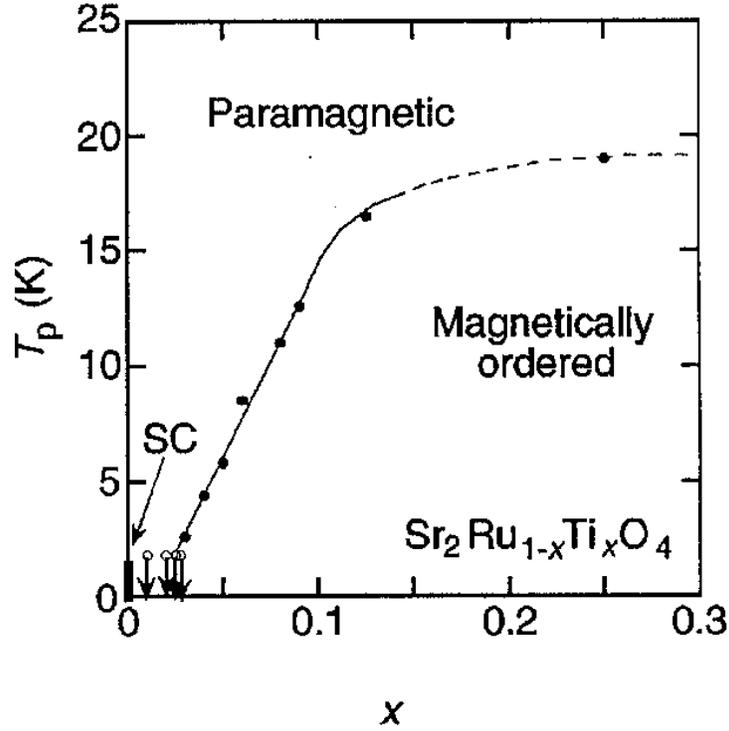} 
\end{center} 
\caption{ Phase diagram of Sr$_2$Ru$_{1-x}$Ti$_x$O$_4$ \protect\cite{kiki}. 
The superconductivity 
is completely suppressed for $x>0.02$ and it becomes magnetically ordered with 
{\bf Q}$_{ic} \sim (2\pi/3,2\pi/3)$ indicating an importance of incommensurate 
antiferromagnetic fluctuations of the $xz$- and $yz$-bands 
for the magnetically-ordered state.
} 
\label{tiphase} 
\end{figure} 
Cooper-pairing. In cuprates the effects induced by magnetic and non-magnetic
impurities are not well understood. For example by doping the non-magnetic Zn
the local magnetic moments in CuO$_2$-plane are induced around these
impurities \cite{zn} which demonstrate Kondo-like behavior. 
On the contrary, the magnetic Ni impurities also show quite puzzling features
which so far have not been understood \cite{vojta}. The complication of
the cuprates are the highly correlated electrons (holes) of the CuO$_2$-plane. 
In this respect it is of high interest to analyze the effect of 
impurities in Sr$_2$RuO$_4$, since it is 
a Fermi-liquid system in the normal state.

Early studies of the impurity effects in Sr$_2$RuO$_4$  revealed some features
reflecting the unconventional nature of superconductivity. For example, the
rapid suppression of $T_c$ by native impurities and defects occurs \cite{mac} as
well as the large enhancement of the residual density of states in the
superconducting state seen in the specific heat \cite{heat} and NMR
measurements \cite{nmr1}. 

Quite recently the first accurate study of the effects induced by the 
magnetic (Ir$^{4+}$) and non-magnetic (Ti$^{4+}$) impurity substitution 
in the RuO$_2$-plane has been performed \cite{kiki}. Here, the observed 
effects are also
quite peculiar. In particular, the substitution of the {\it non-magnetic} 
impurity Ti$^{4+}$ (3$d^{0}$) in Sr$_2$RuO$_4$ induces a local magnetic moment 
with an effective moment $p_{eff} \sim 0.5 \mu_B/$Ti \cite{mina}. The induced 
moment has Ising anisotropy with an easy axis along the $c$-direction. 
The corresponding phase diagram is shown in Fig. \ref{tiphase}.
Furthermore, 
magnetic ordering with glassy behavior appears for $x$(Ti)$>0.025$ in 
Sr$_2$Ru$_{1-x}$Ti$_x$O$_4$ while retaining the metallic conduction along 
the in-plane 
direction. When $x(\mbox{Ti})$ is further increased to $x=0.09$ the elastic neutron 
scattering
measurements detect an incommensurate Bragg peak whose wave 
vector {\bf Q}$_{ic} \sim
(2\pi/3,2\pi/3)$ is close to the position of the inelastic neutron scattering 
peak in pure Sr$_2$RuO$_4$ \cite{brad2}. Most interestingly, in 
the vicinity of a magnetic 
ordering a deviation from a pure Fermi-liquid behavior seen in Sr$_2$RuO$_4$ 
is observed by means of resistivity and transport measurements showing linear and 
logarithmic temperature dependencies, respectively \cite{maeno3}. These results 
indicate that the two-dimensional incommensurate antiferromagnetic spin fluctuations 
arising from the nesting of $xz$- and $yz$-bands become a static spin density wave 
state (SDW) by introducing Ti substitution.
\begin{figure} 
\begin{center} 
\includegraphics[width=0.7\textwidth,angle=0]{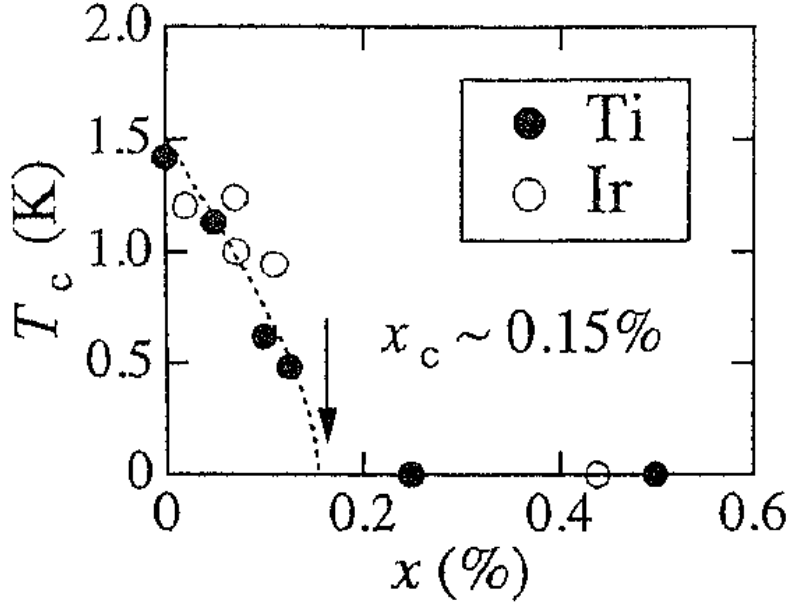} 
\end{center} 
\caption{ Superconducting transition temperature $T_c$ as a function of the 
impurity concentration $x$ (Ti, Ir) \protect\cite{kiki}. 
The superconductivity 
is completely suppressed for $x>0.0015$. The dashed curve shows the fit to 
the Abrikosov-Gor'kov 
pair-breaking function generalized to the case of magnetic and non-magnetic 
impurities in unconventional superconductors \protect\cite{norman}.
} 
\label{kiki1} 
\end{figure} 
On the other hand, the system Sr$_2$Ru$_{1-x}$Ir$_x$O$_4$ in which the 
substitution is magnetic Ir$^{4+}$ (5$d^5$ in the low spin configuration) 
shows 
a weak ferromagnetism for $x(\mbox{Ir}) > 0.3$ \cite{19}. Thus, 
substitution of magnetic and 
non-magnetic impurities in Sr$_2$RuO$_4$ leads to a different ground states.

Despite of these differences both magnetic and 
non-magnetic impurities act similarly 
suppressing superconductivity in Sr$_2$RuO$_4$ for $x>0.0015$ \cite{kiki} as 
shown in Fig. \ref{kiki1}. 
From this behavior it is clear that magnetic and non-magnetic 
impurities act mainly as a potential scatters for superconductivity in 
Sr$_2$RuO$_4$ and that the magnetic scattering does not play any role. 
Although this behavior could be explained in other ways, this observation 
is also consistent with the existence of a spin triplet state. Magnetic 
impurities break singlet Cooper-pairs mainly because of exchange splitting 
of the single particle state. The equal spin paired state would not be 
subject to such an effect. 

More detailed theoretical investigation is still 
needed in order to fully understand the role 
played by the impurities in  Sr$_2$RuO$_4$ and 
illuminate the role played by magnetic fluctuations in the ruthenates.

\section{Conclusions}

In this review we have presented some aspects of superconductivity in 
Sr$_2$RuO$_4$ which have not been reviewed previously in a unified way so far. 
Most importantly, we analyzed the role of electronic 
correlations and the competition between ferromagnetic and incommensurate 
antiferromagnetic spin fluctuations for the formation of unconventional 
triplet superconductivity in Sr$_2$RuO$_4$. We have found that the 
magnetic response is strongly anisotropic and is due to spin-orbit coupling.
As we have shown it plays a crucial role in the formation of 
triplet $p$-wave superconductivity with line nodes between the RuO$_2$-layers.
We further argue that spin-fluctuation-mediated Cooper-pairing scenario in a 
frame of the three-band Hubbard Hamiltonian 
can explain well the symmetry of the superconducting order parameter in 
Sr$_2$RuO$_4$. 
Furthermore, we point out that strong electronic correlations are responsible  
for the rich phase diagram of Ca$_{2-x}$Sr$_x$RuO$_4$ compound upon changing 
the Ca content. 

We have also discussed the open issues in Sr$_2$RuO$_4$, 
like the influence of magnetic and non-magnetic 
impurities on the superconducting and normal state of Sr$_2$RuO$_4$.
It is clear that the physics of triplet superconductivity in Sr$_2$RuO$_4$ 
is still far from being completed and remains to be analyzed more in detail.

We are grateful to M. Sigrist, M. Braden, and K. H. Bennemann 
for stimulating discussions. 
The present work is supported by the INTAS Grant No. 01-654.


\begin{thebibliography}{10}
%
\bibitem{bednorz} J. G. Bednorz and K. A. M\"uller, Z. Phys. B {\bf 64}, 189
  (1986).
%
\bibitem{maeno} Y. Maeno {\it et al.}, Nature {\bf 372}, 532 (1994).
%
\bibitem{mac} A. P. Mackenzie {\it et al.,} Phys. Rev. Lett. {\bf 80}, 
161 (1998).
%
\bibitem{imad} K. Ishida {\it et al.,} Phys. Rev. B {\bf 56}, 505 (1997).
%
\bibitem{manfred} T. M. Rice and M. Sigrist, J. Phys. Condens. Matter {\bf 7}, 
L643 (1995). 
%
\bibitem{sigri} M. Sigrist, D. Agterberg, A. Furusaki, C. Honerkamp,
  K. K. Ng, T. M. Rice, and M. E. Zhitomirsky, Physica C {\bf 317-318},
  134 (1999).
%
\bibitem{ins} Y. Sidis {\it et al.,} Phys. Rev. Lett. {\bf 83}, 3320
(1999).
%
\bibitem{ishida} K. Ishida, H. Mukuda, Y. Kitaoka, Z. Q. Mao, H. Fukuzawa,
and Y. Maeno, Phys. Rev . B \textbf{63}, 060507 (2001).
%
\bibitem{maki} T. Dahm, H. Won, and K. Maki, cond-mat/0006151 (unpublished).
%
\bibitem{maemac} A. P. Mackenzie and Y. Maeno, Rev. Mod. Phys.{\bf 75}, 657 
(2003).
%
\bibitem{noce} C. Noce and M. Cuoco, Phys. Rev. B {\bf 59}, 2659 (1999).
%
\bibitem{oguchi} T. Oguchi, Phys. Rev. B {\bf 51}, 1385 (1995).
%
\bibitem{singh} D. Singh, Phys. Rev. B{\bf 52}, 1358 (1995).
%
\bibitem{liebsch} A. Liebsch and A. Lichtenstein,
Phys. Rev. Lett. {\bf 84}, 1591 (2000).
%
\bibitem{macki} A. P. Mackenzie {\it et al.,} Phys. Rev. Lett. {\bf 76}, 
3786 (1996); Y. Yoshida {\it et al.}, J. Phys. Soc. Jpn. {\bf 68}, 3041 
(1999).
%
\bibitem{singh0} I. I. Mazin and D. Singh, Phys. Rev. B {\bf 56}, 2556 (1997).
%
\bibitem{singh1} I. I. Mazin and D. Singh, Phys. Rev. Lett. 
{\bf 82}, 4324 (1999).
%
\bibitem{imai} T. Imai {\it et al.}, Phys. Rev. Lett. {\bf 81}, 3006 (1998).
%
\bibitem{annett} J. F. Annett, Adv. Phys. {\bf 39}, 83 (1990).
%
\bibitem{60} J. A. Duffy {\it et al.}, Phys. Rev. Lett. {\bf 85}, 5412 (2000).
%
\bibitem{61} G. M. Luke {\it et al.}, Nature {\bf 394}, 558 (1998).
%
\bibitem{scalapino} D. J. Scalapino, Phys. Rep. {\bf 250},
329 (1995); N.~F. Berk and J.~R. Schrieffer,
Phys. Rev. Lett. {\bf 17}, 433 (1966); A. J. Layzer and D. Fay, 
{\it Proc. Int. Low Temp. Conf.},
St. Andrews (LT-11), Academic Press, London, (1968); 
P. W. Anderson and W. F. Brinkmann, in {\it "The 
physics of liquid and solid Helium"}, eds. K. H. Bennemann and
J. B. Ketterson (V. 2, Wiley-Interscience, 1978).
%
\bibitem{ueda} H. Kontani and K. Ueda, Phys. Rev. Lett. {\bf 80},
  5619 (1998).
%
\bibitem{arita} K. Kuroki, M. Ogata, R. Arita, and H. Aoki,
  Phys. Rev. B {\bf 63}, 060506(R) (2001).
%
\bibitem{62} S. Nishizaki, Y. Maeno, and Z. Mao, J. Low Temp. Phys. {\bf 117},
1581 (1999); J. Phys. Soc. Jpn. {\bf 69}, 572 (2000). 
%
\bibitem{64} K. Ishida {\it et al.}, Phys. Rev. Lett. {\bf 84}, 5387 (2000).
%
\bibitem{65} M. Tanatar  {\it et al.}, Phys. Rev. B {\bf 63}, 064505 (2001).
%
\bibitem{67} I. Bonalde {\it et al.}, Phys. Rev. Lett. {\bf 85}, 4775 (2000).
%
\bibitem{ilya1} I. Eremin, D. Manske, C. Joas, and K. H. Bennemann,
Europhys. Lett. {\bf 57}, 447 (2002).
%
\bibitem{rice} M. E. Zhitomirsky and T. M. Rice, Phys. Rev. Lett. {\bf 87},
057001 (2001). 
%
\bibitem{litak} J.F. Annett, G. Litak, B. L. Gyorffy, and K. I. 
Wysokinski, Phys. Rev. B {\bf 66}, 134514 (2002).
%
\bibitem{izawa} K. Izawa, H. Takahashi, H. Yamaguchi, Y. Matsuda, 
M. Suzuki, T. Sasaki, T. Fukase, Y. Yoshida, R. Settai, and Y. Onuki, 
Phys. Rev. Lett. {\bf 86}, 2653 (2001).
%
\bibitem{hase} Y. Hasegawa and M. Yakiyama, J. Phys. Soc. Jpn. {\bf 72}, 
1318 (2003).
%
\bibitem{yana} Y. Yanase and M. Ogata, J. Phys. Soc. Jpn. {\bf 72}, 
673 (2003). 
%
\bibitem{shigeru} S. Koikegami, Y. Yoshida, and T. Yanagisawa, 
Phys. Rev. B {\bf 67}, 134517 (2003).
%
\bibitem{ng2} K. K. Ng and M. Sigrist, Europhys. Lett. {\bf 49},
473 (2000). 
%
\bibitem{saw} T. Mizokawa, L. H. Tjeng, G. A. Sawatzky, G. Ghiringhelli, 
O. Tjernberg, N. B. Brookes, H. Fukazawa, S. Nakatsuji, and Y. Maeno, 
Phys. Rev. Lett. {\bf 87}, 077202 (2001).
%
\bibitem{ilya2} I. Eremin, D. Manske, and K. H. Bennemann, Phys. Rev. B 
{\bf 65}, 220502(R) (2002); I. Eremin, D. Manske, J.-R. Tarento, and 
K. H. Bennemann, J. Supercond. {\bf 15}, 447 (2002). 
%
\bibitem{ng1} K. K. Ng and M. Sigrist, J. Phys. Soc. Jpn. {\bf 69}, 3764 
(2000). 
%
\bibitem{brade} M. Braden, P. Steffens, Y. Sidis, J. Kulda, P. Bourges, 
S. Hayden, N. Kikugawa, and Y. Maeno, cond-mat/0307662 (unpublished).
%
\bibitem{maenosec} K. Deguchi, M. Tanatar, Z. Mao, T. Ishiguro, and 
Y. Maeno, J. Phys. Soc. Jpn. {\bf 71}, 2839 (2002).
%
\bibitem{physto} Y. Maeno, T. M. Rice, and M. Sigrist, Phys. Today,
  No. 1, 42 (2001).
%
\bibitem{ovchin} E. V. Kuz'min, S. G. Ovchinnikov, and I. O. Baklanov,
Phys. Rev. B {\bf 61}, 15392 (2000).
%
\bibitem{zhang} F. C. Zhang and T. M. Rice, Phys. Rev. B {\bf 37}, (1987).
%
\bibitem{khb} K. H. Bennemann and J. Ketterson (eds.), Physics of
Superconductors (Springer-Verlag, Berlin, 2003).
%
\bibitem{naka} S. Nakatsuji and Y. Maeno, Phys. Rev. Lett. {\bf 84}, 2666
(2000); Phys. Rev. B {\bf 62}, 6458 (2000).
%
\bibitem{friedt} O. Friedt {\it et al.}, Phys. Rev. B {\bf 63}, 174432 (2001).
%
\bibitem{feng} Z. Fang and K. Terakura, Phys. Rev. B {\bf 64}, 020509 (2001).
%
\bibitem{ovchin1} S. G. Ovchinnikov, in: Ruthenate and Rutheno-Cuprate
Materials, Lecture Notes in Physics (Springer, Berlin, 2002).
%
\bibitem{mizo} T. Mizokawa {\it et al.}, Phys. Rev. Lett. {\bf 87}, 077201
(2001). 
%
\bibitem{naka1} F. Nakamura, T. Goko, M. Ito, T. Fujita, S. Nakatsuji, 
H. Fukazawa, Y. Maeno, P. Alireza, D. Forsythe, and S. R. Julian,  
Phys. Rev. B {\bf 65}, 220402(R) (2002).
%
\bibitem{anisim} V. I. Anisimov, I. A. Nekrasov, D. E. Kondakov, T. M. Rice, 
and M. Sigrist, Eur. Phys. Journal B {\bf 25}, 191 (2002).
%
\bibitem{zn} A. V. Mahajan et al., Phys. Rev. Lett. {\bf 72}, 3100 (1994).
%
\bibitem{vojta} A. Polkovnikov, S. Sachdev, and M. Vojta, 
Phys. Rev. Lett. {\bf 86}, 296 (2001).
%
\bibitem{heat} S. Nishizaki, Y. Maeno, and Z. Q. Mao, J. Low Temp. Phys. {\bf
117}, 1581 (1999).
%
\bibitem{nmr1} K. Ishida, H. Mukuda, Y. Kitaoka, Z. Q. Mao, H. Fukazawa, and
Y. Maeno, Phys. Rev. Lett. {\bf 84}, 5387 (2000).
%
\bibitem{kiki} N. Kikigawa, A. P. Mackenzie, and Y. Maeno,
J. Phys. Soc. Jpn. {\bf 72}, 237 (2003).
%
\bibitem{mina} M. Minakata and Y. Maeno, Phys. Rev. B {\bf 63}, 180504(R) (2001).
%
\bibitem{brad2} M. Braden {\it et al.,} Phys. Rev. Lett. {\bf 88}, 197002 (2002).
%
\bibitem{maeno3} N. Kikigawa and Y. Maeno, Phys. Rev. Lett. {\bf 89}, 117001 (2002).
%
\bibitem{19} R. J. Cava, B. Batlogg, K. Kiyono, and H. Takagi, Phys. Rev. B {\bf 49} 
11890 (1994).
%
\bibitem{norman} R.J. Radtke, K. Levin, H.-B. Sch\"uttler, and M. Norman, Phys. Rev.
B {\bf 48}, 653 (1993). 
%
\end{thebibliography}
\end{document}